\documentclass[twocolumn]{revtex4}

\usepackage{ORI_Group_style}
\usepackage[normalem]{ulem}
\usepackage{float}

\begin{document}

\title{On-Chip Quantum Interference of a Superconducting Microsphere}

\author{H. Pino}
\author{J. Prat-Camps}
\author{K. Sinha}
\author{B. Prasanna Venkatesh}
\author{O. Romero-Isart}

\email{oriol.romero-isart@uibk.ac.at}
\affiliation{Institute for Quantum Optics and Quantum Information of the
Austrian Academy of Sciences, A-6020 Innsbruck, Austria.}

\affiliation{Institute for Theoretical Physics, University of Innsbruck, A-6020 Innsbruck, Austria.}

\begin{abstract}

We propose and analyze an all-magnetic scheme to perform a Young's double slit experiment with a  micron-sized superconducting sphere of mass $ \gtrsim 10^{13}$ amu. We show that its center of mass could be prepared in a spatial quantum superposition state with an extent of the order of half a micrometer. The scheme is based on magnetically levitating the sphere above a superconducting chip and letting it skate through a static magnetic potential landscape where it interacts for short intervals with quantum circuits. In this way, a protocol for fast quantum interferometry using quantum magnetomechanics is passively implemented. Such a table-top earth-based quantum experiment would operate in a parameter regime where gravitational energy scales become relevant. In particular, we show that the faint parameter-free gravitationally-induced decoherence collapse model, proposed by Di\'osi and Penrose, could be unambiguously falsified. 
\end{abstract}

\maketitle

\section{Introduction}

\begin{figure}[t]
\centering
\includegraphics[width=  \columnwidth]{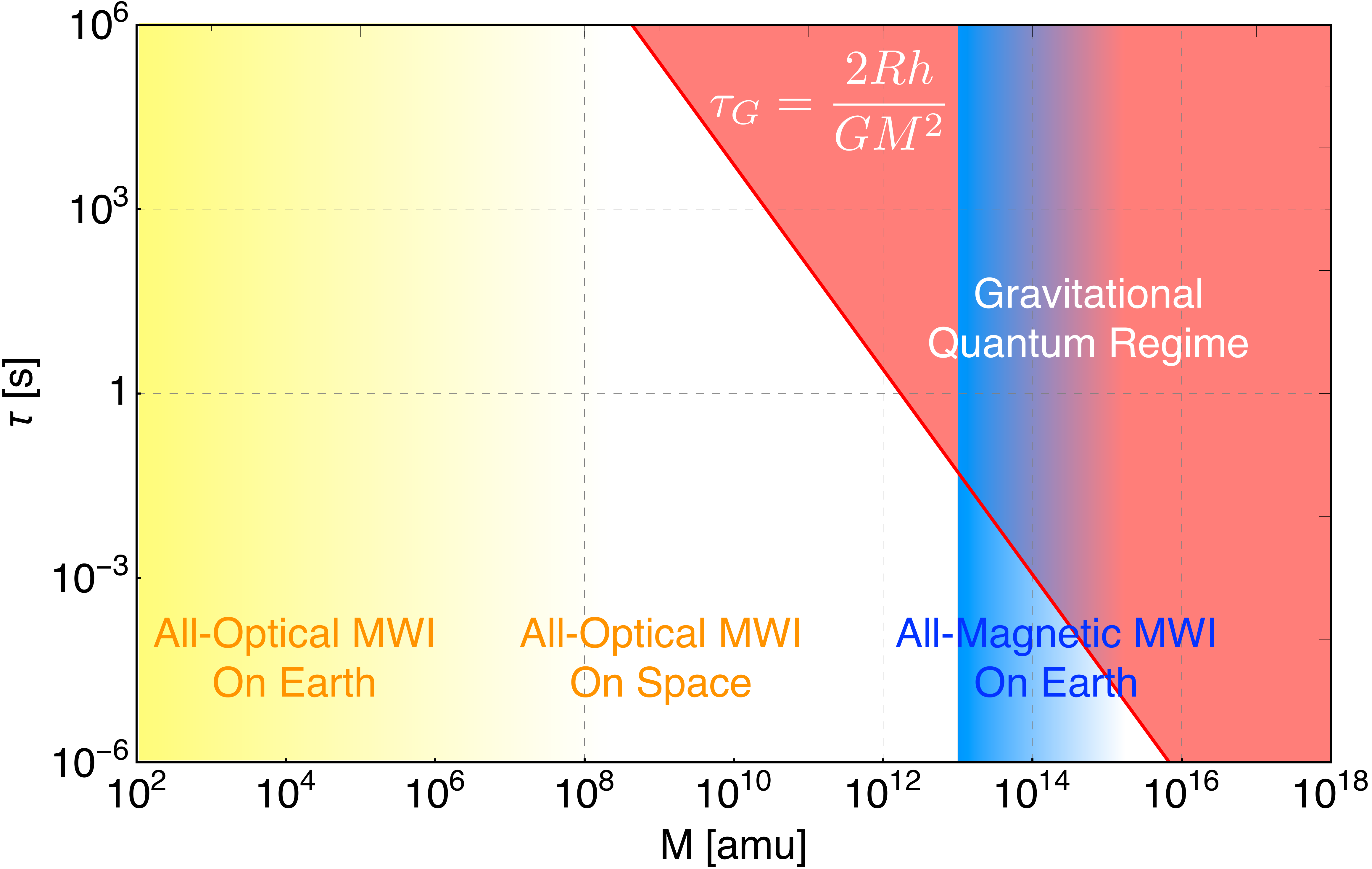}
\caption{{Time scale $\tau_G \equiv 2 R h/(GM^2)$ is plotted as a function of the mass $M$ for a sphere of radius $R$ and a typical mass density of a metallic solid object $10^4~\text{Kg}/\text{m}^3$.}}
\label{Fig1}
\end{figure}

Preparing a massive object in a spatial quantum superposition over distances comparable to its size is a tantalizing possibility. This has been achieved in seminal experiments on matter-wave interferometry of molecules~\cite{Arndt1999,Hornberger2012,Cronin2009,Arndt2014}, which hold a mass record of $10^{4}$ amu~\cite{Eibenberger2013}. It would be particularly intriguing to prepare large quantum superpositions of even larger masses such that one could enter into the {\em gravitational quantum regime} (GQR), which we define as follows. For a solid sphere of radius $R$ and mass $M$, we define the timescale $\tau_G \equiv 2 R h/(GM^2)$, where $h$ is the Planck's constant and $G$ the Newton's gravitational constant. $\tau_G$ has two interpretations: (i) it is the conjectured lifetime of a  quantum superposition state of a single sphere delocalized over a distance $R$, according to the parameter-free~\cite{Footnote1,Diosi2007,ORI2011b} gravitationally-induced decoherence collapse model proposed by Di\'osi and Penrose~\cite{Diosi1984,Penrose1996}, (ii) $h/\tau_G$ is the kinetic energy of two spheres equivalent to the gravitational interaction energy of two point particles of the same mass separated by a distance $2 R$, a situation that could be used to measure $G$~\cite{Schmoele2016}. Moreover, $\tau_G$ is  related   to the much longer timescale $\tau_G R^2/x_0^2$ required to entangle the center-of-mass motion of two spheres spatially localized within a length scale $x_0$ and separated by a distance $2R$ using the gravitational interaction~\cite{Zeh2011,Anastopoulos2015}.
For a given mass density, one can plot $\tau_G$ as a function of $M$, see \figref{Fig1}. We define the GQR as the regime in which $\tau_G < \tau$, where $\tau$ is a characteristic time scale of the quantum system for each of the two interpretations of $\tau_G$. In particular: in (i) $\tau$ is the coherence time of the superposition taking into account standard sources of decoherence, in (ii) $2 \pi/\tau$ could be the trap frequency of a harmonic potential where two spheres are cooled near to their respective ground states. 
Choosing $\tau$ to be of the order of $1$ second, which is an ambitious timescale for quantum experiments, one sees in \figref{Fig1} that $\tau_G < 1~\text{s}$ pertains to masses $> 10^{12}~\text{amu}$. The goal of this article is to propose an earth-based experimental scenario that allows one to access the GQR.

{Preparing a large mass ($> 10^{12}~\text{amu}$) in a large quantum superposition (comparable to its size) is an ambitious challenge that can be addressed via two different approaches: either the {\em bottom-up approach}, consisting in gradually increasing the mass of the object that is prepared in a large superposition, or the {\em top-down approach}, that gradually increases the delocalization distance of a large mass that is brought into the quantum regime. Matter-wave interferometry has traditionally followed the bottom-up approach. Starting in the 30's with the quantum interference of single electrons, the mass of the objects that can be interfered has remarkably increased up to molecules of $10^{4}$~amu~\cite{Eibenberger2013}. The top-down approach is more recent and has been triggered by the field of quantum nano- and micromechanics~\cite{revopt}. Recent achievements in this field include bringing objects with masses ranging from $10^{11}$ amu to $10^{13}$ amu to the quantum regime by cooling their associated mechanical degree of freedom to its quantum ground state~\cite{oconell10,teufel11,chan11,revopt}. However, while the mass is definitely large, the delocalization distance of the ground state in such systems is typically smaller than the size of a single atom. Such tiny delocalization distances prevent current quantum micromechanical oscillators to enter into the GQR, specially in the context of (i). Levitated mechanical oscillators, in particular optically levitated dielectric nanospheres in high finesse optical cavities~\cite{ORI2010,Chang2010, Barker2010,Kiesel2013,Asenbaum2013,Millen2015,Gieseler2012}, have been suggested as an ideal nanomechanical system to implement the top-down approach by releasing the mechanically cooled nanosphere from the trap to increase the delocalization distance~\cite{ORI2011,ORI2011b}. However, as shown in Fig.~\ref{Fig1}, these all-optical hybrid schemes~\cite{ORI2011,ORI2011b,Bateman2014} using the best of matter-wave interferometry and quantum optomechanics seem, even in the most ambitious proposals requiring a space environment~\cite{MAQRO,Kaltenbaek2015}, to be limited to masses between $10^8$ and $10^{11}$ amu, putting them still away from the GQR.}

Here we propose to attain spatially large quantum superpositions of masses $\gtrsim 10^{13}$ amu, well inside the GQR, by abandoning the use of lasers and using instead an {\em all-magnetic} on-chip architecture. This approach combines the following salient features: (i) cryogenic temperatures both for environment and the massive particle to minimize decoherence due to emission, scattering, and absorption of black-body radiation, (ii) the use of static magnetic potentials created by persistent currents to diamagnetically levitate the sphere~\cite{RomeroIsart2012} without creating decoherence (space environment is not required) as well as to exponentially speed-up quantum dynamics by using inverted potentials~\cite{RomeroIsart2016}, and (iii) coupling the position of the sphere to quantum circuits on an integrated on-chip configuration to cool the center-of-mass motion to the ground state, to make a double-slit smaller than the size of the sphere by measuring the squared center-of-mass position~\cite{ORI2011}, and to measure the interference pattern downstream. These tools permit, in principle, to make an all-magnetic Young's double slit experiment, with a single run-time of less than a second, on-earth, with a superconducting microsphere of radius $1~\mu$m, mass $10^{13}$~amu, and with a slit separation of half a micrometer. As discussed in~\cite{RomeroIsart2016}, this experimental proposal makes an optimal use of inverted potentials.

  The article is organized as follows. In \secref{Sec:Challenges} we review the main challenges that a proposal aiming at preparing large quantum superpositions of a massive object has to overcome. The discussion of these challenges aims at motivating both the protocol and the implementation of our proposal. In \secref{Sec:QIProtocol} we describe the quantum micromechanical interferometer protocol as well as how to analytically calculate all the steps of the protocol taking into account decoherence. This enables us to analytically obtain the interference pattern as a function of all the physical parameters of the protocol and environmental decoherence. In \secref{Sec:QIImplementation} we propose and analyze an all-magnetic on-chip implementation of the quantum micromechanical interferometer protocol for a superconducting microsphere. In \secref{Sec:CaseStudy} we analyze a case study with a set of experimental parameters whose complete list is given in the Appendix. 
We draw our conclusions and provide some final remarks in \secref{Sec:Conclusions}.

\section{Challenges} \label{Sec:Challenges}

Let us discuss three fundamental challenges that will need to be overcome by proposals aiming to prepare large superpositions of a massive object in the GQR: unavoidable decoherence, slow quantum dynamics, and the double-slit. These challenges are discussed in a general fashion without considering any particular implementation. They will serve as a motivation to introduce the quantum micromechanical interferometer protocol in~\secref{Sec:QIProtocol} and the proposed experimental implementation in~\secref{Sec:QIImplementation}.

\subsection{Unavoidable decoherence}

Consider a sphere of radius $R$ whose center-of-mass position along the $x$-axis is described by the quantum state $\hat \rho$. Most sources of decoherence of the center-of-mass position are of the position localization (PLD) type (see, for example~\cite{ORI2011b,Joos2003,Schlosshauer2007}). PLD makes the off-diagonal terms of the density matrix in the position basis decay exponentially in time. In the so called long wavelength (LW) limit, where decoherence events provide only partial which-path information, it can be shown that $\bra{x} \hat \rho(t) \ket{x'} \propto \exp[- \Lambda (x-x')^2 t] \bra{x} \hat \rho(0) \ket{x'}$, where $\Lambda$ is the localization parameter. In this regime decoherence is modeled by the master equation 
\be 
\dot \rho(t) = \frac{1}{\im \hbar} \coms{ \Hop}{\hat \rho} - \Lambda \coms{\xop}{\coms{\xop}{\hat \rho}} 
\label{eq:PLdecoh},
\ee
where $\Hop$ is the Hamiltonian of the system. In the short wavelength (SW) limit, where decoherence events provide a full which-path information, one has that $\bra{x} \hat \rho(t) \ket{x'} \propto \exp[- \gamma t] \bra{x} \hat \rho(0) \ket{x'}$. 

As discussed in \cite{ORI2011b}, most of the paradigmatic collapse models predict, at the level of density matrices, a PLD. In particular, the effect of the parameter-free gravitationally induced collapse model, conjectured by Penrose and Di\'osi~\cite{Diosi1984,Penrose1996}, can be described with the LW limit of the PLD with a localization parameter given by~\cite{ORI2011b}
\be
\Lambda_G = \frac{G M^2}{2 R^3 \hbar}.\label{eq:DPlam}
\ee
Here, $G$ is the Newton's gravitational constant. {Note that $\Lambda_G$ is parameter-free, namely it does not depend on any phenomenological parameter since a homogeneous mass density is assumed. Such choice provides a lower bound on the gravitationally-induced decoherence rate. Hence, by falsifying this case, all other parameter-dependent forms of gravitationally-induced decoherence would also be falsified, see~\cite{Diosi2007,ORI2011b} for further details}. Combining Eqs.~\ref{eq:PLdecoh} and \ref{eq:DPlam}, we can immediately see that the correlations in a spatial quantum superposition of extent $R$ decay as $\bra{R/2} \hat \rho(t) \ket{-R/2} \propto \exp[- 2 \pi t/\tau_G] \bra{R/2} \hat \rho(0) \ket{-R/2}$, where $\tau_G^{-1}=\Lambda_G R^2/(2 \pi)$ is the timescale used in \figref{Fig1} to define the GQR. 

In order to falsify such an exotic source of decoherence, which predicts the breakdown of standard quantum mechanics, one has to account for the standard sources of decoherence that are expected in an implementation. In this section, we review the following inevitable sources of decoherence: scattering of gas molecules, emission, absorption, and scattering of black-body radiation, and stochastic forces due to vibrations.

Decoherence due to scattering of gas molecules, with a thermal wavelength typically smaller than the superposition size of interest, is described by PLD in the SW limit~\cite{ORI2011b}. The decoherence rate is given by
\be 
\gamma_a = \frac{16 \pi \sqrt{2 \pi}}{\sqrt{3}}  \frac{PR^2}{\sqrt{3 m_a K_b T_e}},
\ee 
where $P$ is the environmental pressure, $T_e$  the environmental temperature, $K_b$ the Boltzmann constant, and $m_a$  the mass of a gas molecule. In order to preserve coherence, the quantum micromechanical interferometer should be implemented in a time $T \ll \gamma_a^{-1}$, namely, it is required that the sphere does not scatter any gas molecules during each run of the protocol. For a sphere of $R=1~\mu\text{m}$ in an optimal cryogenic environment with $T_e=50~\text{mK}$ and $P=10^{-17}~\text{mbar}$, the coherence time is given by $\gamma_a^{-1} \approx 1.6~\text{s}$. Hence, a single run of the quantum micromechanical interferometer has to be performed in a timescale of the order of a second or less. Combined with the slow quantum dynamics of large masses, as we discuss below, this poses a challenge. We remark that pressures as low as $10^{-17}~\text{mbar}$ can be achieved in cryogenic environments. This was shown, for example, in the experiments with antiproton lifetimes by Gabrielse \textit{et al}~\cite{Gabrielse90, GabrielseRev}. Note that $\gamma_a^{-1}$ is an upper-bound for the time scale $\tau$ used to define the GQR, especially regarding interpretation (i).

Decoherence due to scattering, absorption, and emission of black-body radiation is described by PLD in the LW limit, since the wavelength of thermal photons is typically much larger than the superposition size. The localization parameter is given by $\Lambda_{bb}=\Lambda^s_{bb}+\Lambda^e_{bb}+\Lambda^a_{bb}$, where~\cite{Joos2003,Schlosshauer2007, ORI2011b,MAQRO}
\be
\Lambda^s_{bb} \approx \frac{8! 8 \zeta(9)}{9 \pi} c R^6 \pare{\frac{K_b T_e}{\hbar c}}^9  \text{Re} \spare{\frac{\epsilon(\w_\text{th}) - 1}{\epsilon(\w_\text{th}) +2 }}^2
\ee
is the contribution due to scattering and
\be \label{eq:LBBe}
\Lambda^{e(a)}_{bb} \approx \frac{16 \pi^5}{189} c R^3 \pare{\frac{K_b T_{i(e)}}{\hbar c}}^6 \text{Im} \spare{\frac{\epsilon(\w_\text{th}) - 1}{\epsilon(\w_\text{th}) +2 }}
\ee
are the contributions due to emission (e) and absorption (a) of thermal photons. The last two contributions are the same when the internal bulk temperature $T_i$ of the sphere is the same as the environmental $T_e$. These localization parameters depend on the dielectric permittivity $\epsilon$ of the sphere at the thermal frequency $\omega_{\text{th}}$. For a sphere of $R=1~\mu\text{m}$ and density $10^4~\text{Kg}/\text{m}^3$, one can show that
\bea  
\frac{\Lambda^s_{bb}}{\Lambda_G} &\approx& 10^{-15} \times \pare{\frac{T_e}{1~\text{K}}}^9\text{Re}\spare{\frac{\epsilon(\w_\text{th}) - 1}{\epsilon(\w_\text{th}) +2 }}^2 \\
\frac{\Lambda^{e(a)}_{bb}}{\Lambda_G} &\approx& 10^{-7} \times \pare{\frac{T_{i(e)}}{1~\text{K}}}^6 \text{Im} \spare{\frac{\epsilon(\w_\text{th}) - 1}{\epsilon(\w_\text{th}) +2 }}
\eea
Hence, in order to minimize the effect of decoherence due to black-body radiation, such that gravitationally-induced decoeherence would dominate, the quantum micromechanical interferometer should be implemented in a cryogenic environment.

Any experiment will suffer from unavoidable vibrational noise. Let us consider vibrations along the $x$-axis described by the fluctuating variable $\xi_x(t)$, which has length units and fulfills $\avg{\xi_x(t)}=0$ and $\avg{\xi_x(t) \xi_x(t')} \neq 0$. The vibrational noise is characterized by the power spectral density
\be
S_{xx}(\w) = \intall \text{d} \tau \avg{ \xi_x (t+\tau) \xi_x (t)} e^{\im \w \tau},
\ee
which has units of $\text{m}^2/\text{Hz}$. For a given external potential $V(\xop)$ that depends on the $x$-coordinate of the center-of-mass of the sphere, the vibrational noise leads to a time-dependent stochastic force given by $f(t) \equiv - V'[\xi_x(t)]$. The Hamiltonian  of the system  can then be written as 
$\Hop(t) = \Hop + f(t) \xop$. For a harmonic potential of the form $V(\xop)=M \w_p^2 \xop^2/2$ with the stochastic force due to vibrations given by $f(t) = - M \w_p^2 \xi_x(t)$, one can show~\cite{Henkel1999,Breuer2002} that by averaging the stochastic force, one obtains a master equation for the dynamics. For the particular case of $S_{xx}(\w_p) = S_{xx}(-\w_p)$, this master equation is of the PLD form in the LW limit with a localization parameter given by
\be
\Lambda_{v} = \frac{M^2 \w_p^4}{2 \hbar^2} S_{xx}(\w_p).
\ee
The decoherence due to vibrations scales quadratically (quartically) with the mass (with the trap frequency) of the sphere since the stochastic force is proportional to $M \w_p^2$ for a harmonic potential. As expected, the stochastic force is zero when the particle is not in an external potential. In order to falsify gravitationally-induced decoherence in the presence of an external harmonic potential, the decoherence induced by such fluctuations has to satisfy $\Lambda_v/\Lambda_G =R^3 \w_p^4 S_{xx}(\w_p)/(G \hbar) < 1$, which becomes increasingly challenging with the size of the sphere. In particular, for a sphere of $R=1~\mu\text{m}$ one requires $\sqrt{S_{xx}(\w_p)} < [(2\pi\times 1~\text{Hz})/\w_p]^2 \times 10^{-15}~\text{m}/\sqrt{\text{Hz}}$. Thus, in order to suppress the effects of vibration-induced decoherence, the implementation of the quantum micromechanical interferometer necessitates a minimal use of external potentials in addition to substantial vibration isolation.

\subsection{Slow quantum dynamics}

In any quantum micromechanical interferometer scheme there are two important dynamical processes that require some time: the generation of a large enough coherence length, to be defined below, and the generation of an interference pattern with sufficient visibility, namely that the separation between the interference fringes is larger than the resolution with which the position can be measured.

In order to define the coherence length, let us first recall some properties of an important class of states -- Gaussian states -- that can be used to describe the quantum state of the sphere along the $x$-axis. A quantum state $\hat \rho$ with $\tr [ \xop \hat \rho] = \tr [ \pop \hat \rho] =0$ is called Gaussian if it is completely determined by the following three real parameters $v_x \equiv \tr [  \xop^2 \hat \rho]$, $v_p \equiv \tr [ \pop^2 \hat \rho]$, and $c \equiv \tr [  (\xop \pop + \pop \xop) \hat \rho]/2$ \cite{Steck2014}.
Indeed, the density matrix of a Gaussian state in the position basis is given by
\be\label{Gausstate}
\bra{x} \hat \rho \ket{x'} = \sqrt{\frac{a_1+ a_1^* +a_2}{\pi}}e^{ -a_1 x^2 - a_1^*  x'^2 - a_2 x x'},
\ee
where
\be
\begin{split}
v_x a_1 &\equiv \frac{\mathcal{P}^2+1}{8 \mathcal{P}^2} - \im \frac{c}{2 \hbar}, \\
v_x a_2 &\equiv \frac{\mathcal{P}^2-1}{4 \mathcal{P}^2}.
\end{split}
\ee
Here we have used that the purity $\mathcal{P} \equiv \tr [\hat \rho^2]$ of a Gaussian state is given by
 $\mathcal{P} = \hbar [4v_x v_p - 4c^2]^{-1/2}$.
Note that the Heisenberg uncertainty relation for a Gaussian state reads as $ v_x v_p - c^2 \ge \hbar^2/4$.

The coherence length $\xi$ for a Gaussian state is defined as
$\bra{x/2} \hat \rho \ket{-x/2} =  \exp [- x^2/\xi^2]/\sqrt{2 \pi v_x}$,
and is given by the compact expression
\be
\xi\equiv \mathcal{P} \sqrt{8 v_x}.
\ee
The coherence length scales with the position variance as $\sqrt{v_x}$ with a scaling factor given by the purity of the state. For a Gaussian state with a coherence length $\xi$ incident on a double slit of separation $d$, one will observe a coherent superposition resulting in interference fringes for $\xi \gg d$ and an incoherent mixture for $\xi \ll d$.

It is straightforward to show that during free expansion of the wavefunction without decoherence, or in other words, under unitary evolution with $\Hop=\pop^2/(2M)$, the coherence length grows at a speed given by
\be
 \lim_{t \rightarrow \infty}\dot \xi(t) =  \mathcal{P} \frac{\sqrt{8 v_p(0)}}{M}.
\ee
Starting out in the ground state of a harmonic potential with frequency $\omega$, such that $\mathcal{P}=1$ and $v_p(0)=\hbar \omega M/2$, then $\lim_{t \rightarrow \infty}\dot \xi(t) = (4\hbar \w/M)^{1/2}$. Note that  the larger the mass the slower the coherence length grows. For a sphere of $R=1~\mu\text{m}$ and mass density $10^4 ~\text{Kg}/\text{m}^3$ initially in a harmonic trap of frequency $\w=2\pi \times 10^5~\text{Hz}$, the speed at which the coherence length grows is $80~\text{nm}/s$. Since we would like to have $\xi$ of the order of $R$, this speed in free expansion poses a challenge for propagation times of the order of a second, as required to prevent decoherence due to scattering of gas molecules.

Even more challenging is the time required to generate sufficiently separated fringes. It is known in matter-wave interferometery that after the Gaussian state passes through a double slit of separation $d \ll \xi$, one obtains a coherent superposition which after free expansion downstream generates, in the absence of decoherence, an interference pattern with a fringe spacing
\be 
x_f = \frac{2 \pi \hbar t}{M d}.
\ee 
The fringe separation speed is thus given by $\dot x_f=2 \pi\hbar/(Md)$. For a sphere of $R=1~\mu\text{m}$, mass density $10^4~\text{Kg}/\text{m}^3$, and assuming a double slit of separation $d=2R$, the speed is given by $\dot x_f \approx 10^{-5} \text{nm}/\text{s}$ which is very small. For example, assuming a position measurement resolution of $1~\text{nm}$, one would require $10^5~\text{s}$ to obtain resolvable fringes. This definitely poses a very important challenge.

\subsection{Double slit}

The final challenge is to prepare a quantum superposition of the center-of-mass position of the sphere, which should ideally be done without any decoherence. This is a non-Gaussian operation that transforms a Gaussian state with a positive Wigner function into a non-Gaussian state with a negative Wigner function. The preparation of non-Gaussian states has been a major goal and a rather formidable challenge in quantum nanomechanics~\cite{revopt}. A double slit in a Young's experiment is a beautiful example of how this can be done. However, in a double slit one is typically forced to prepare spatial superpositions which are larger than the size of the object, namely $d>2R$. For large masses this condition might be too demanding as it could compromise the conditions required to meet the challenges discussed before. For instance the larger the superposition the longer it takes for visible fringes to be generated. Hence, it might be advantageous to prepare superpositions with an extent not necessarily larger than the diameter of the sphere. Implementing a double slit with a slit separation smaller than the size of the sphere is a seemingly naive challenge, but an important one to meet when taking into account the constraints posed by the slow quantum dynamics and the unavoidable sources of decoherence.

\section{quantum micromechanical interferometer: protocol} \label{Sec:QIProtocol}

\begin{figure*}[t]
\centering
\includegraphics[width=2 \columnwidth]{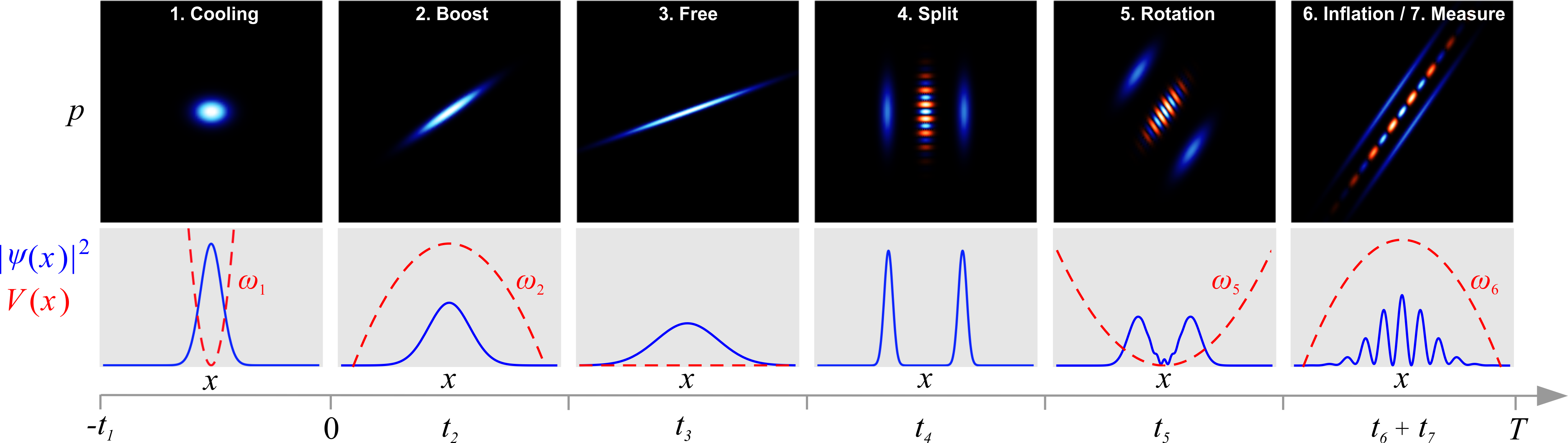}
\caption{Schematic representation of the 7 steps of the quantum micromechanical interferometer protocol. The upper panels show a contour plot of the Wigner function, where the vertical (horizontal) axis represents $p$ ($x$), and the blue (orange) are positive (negative) values. The lower panels show the potential (red dashed line) along the $x$-axis, and the functional form of the probability position distribution along the $x$-axis (solid blue line). The timeline of the protocol is also included. The plotted functions of the different steps are not scaled and arbitrary units are shown, only the functional form has relevant information.}
\label{Fig2}
\end{figure*}

In this section, we propose and analyze a quantum micromechanical interferometer protocol for a massive object. The protocol will be designed in a manner such that two of the challenges discussed in \secref{Sec:Challenges}, namely that of slow quantum dynamics and of the double slit, are addressed. A detailed implementation of the protocol for the specific case of a superconducting sphere is proposed and analysed in \secref{Sec:QIImplementation}. This implementation is designed to overcome the  unavoidable decoherence challenge. Thus in this section we keep the discussion general without reference to any particular implementation. 

The quantum micromechanical interferometer protocol is divided into 7 steps (see~\figref{Fig2});
 \begin{enumerate}
     \item {\bf Cooling}. Cooling the center-of-mass motion in a harmonic trap of frequency $\omega_1$ for a duration $t_1$ to a final phonon occupation number $\bar n$.
     \item {\bf Boost}. Evolution of the center-of-mass motion in an inverted harmonic potential of frequency $\omega_2$ for a time $t_2$ which exponentially boosts the sphere's kinetic energy.
     \item {\bf Free}. Free evolution for a time $t_3$ to delocalize the center-of-mass over large distances.
     \item {\bf Split}. A squared position measurement for a time $t_4$ to prepare a quantum superposition state.
     \item {\bf Rotation}. A short evolution for a duration $t_5$ in a harmonic trap of frequency $\w_5$ to give opposite momentum to the two wave packets in the superposition.
     \item {\bf Inflation}. Evolution in an inverted quadratic potential of frequency $\w_6$ for a duration $t_6$ to exponentially generate the interference fringes.
     \item {\bf Measure}. A squared position measurement of time $t_7$ to unveil the interference pattern.
 \end{enumerate}
  The total time of the protocol after cooling is thus given by $T=\sum_{n=2}^7 t_n$. The position of the center-of-mass of the sphere is given by $\rr = (\xop,y,z)$, where $\xop$ is the position of the center of along the direction in which the quantum superposition is prepared. The motion along the three axes is assumed to be pre-cooled via  feedback cooling. While the motion along the $x$-axis will be further cooled to the ground state via cavity cooling, the other degrees of freedom are assumed to remain in the thermal classical state obtained with feedback cooling. 
  
  Steps 1 to 3 of the protocol are used to create a large and coherent matter-wave of the center of mass of the sphere along the $x$-axis. These are the steps used to generate an observable difference between the state of the sphere when one includes the existence of gravitationally-induced decoherence and when one does not. In particular, the coherence length $\xi$ after step 3, takes a smaller value $\xi_G<\xi$ when gravitationally-induced decoherence is taken into account. In order to falsify gravitationally-induced decoherence we aim at operating the micromechanical interferometer in a regime where $\xi_G \ll d < \xi$, for a given slit separation $d$. This implies that the double slit prepares a coherent superposition in the absence of gravitationally-induced decoherence, and an incoherent mixture otherwise. Steps 4 to 7 consist of performing the double slit operation and unveiling the interference fringes for the case in which the coherent superposition has been prepared. These steps have to be done unitarily such that if one prepares a coherent superposition the visibility of the interference fringes is not washed out. Thus, we emphasize that steps 1 to 3 are used to create the difference between the state with and without the existence of gravitationally-induced decoherence, and steps 4 to 7 are used to unveil this difference. 
  
  In the following, we show how the position probability distribution along the $x$-axis after step 7, namely $P(x)=\bra{x} \hat \rho(T) \ket{x}$, can be analytically calculated taking into account PLD in the LW limit and all the parameters describing the protocol. We divide the discussion into the separate steps of the protocol.

\subsection{Step 1: Cooling}

In this step, which takes place from $t=-t_1$ to $t=0$, the center-of-mass motion of the microsphere along the $x$-axis is cooled using a linear quantum nanomechanical coupling to a cavity mode~\cite{WilsonRae07,Marquardt07,Genes2008,WilsonRae08}. The particle is confined in a harmonic trap of frequency $\w_1$, such that its zero-point motion length is given by $\sigma_1 = \sqrt{\hbar/(2 M \w_1)}$. The total Hamiltonian describing the coupling of the particle to a driven cavity mode of frequency $\w_{c_1}$ and creation (annihilation) mode operators $\adop$ ($\aop$) is given by 
\be
\begin{split}  \label{coolham}
\frac{H_1}{\hbar} =& - \Delta' \adop \aop +  \w_1 \bdop \bop + g_l\adop \aop \left(\bop+\bdop\right) \\ & - \im  E_1 \left( \aop - \adop \right).
\end{split}
\ee
The above Hamiltonian is given in a rotating frame with the drive frequency $\w_{l_1}$. Here $\tilde x=\xop/\sigma_1 = (\bop+\bdop)$ with $[\bop,\bdop]=1$, $\Delta'=\omega_{l_1} - \omega_{c_1}$, $E_1$  is the strength of the coherent drive that couples to the cavity mode, and $g_l$ is the linear quantum nanomechanical coupling. We linearize the system by making the change of variables  $\aop \rightarrow \alpha_1 + \aop$ and $\bop \rightarrow \beta_1 + \bop$, and choosing $\alpha_1$ and $\beta_1$ such that 
\aleqn{
\im E_1 - \left( \Delta'+i\kappa_1 \right) \alpha_1 + g_l\alpha_1 \left(\beta_1+\beta^*_1\right) &= 0\\
\omega_1 \beta_1 + g_l \vert \alpha_1 \vert^2 &=0.
}
Assuming that $g_l \abs{\alpha_1}/(2\kappa_1) \ll 1$ and neglecting the decoherence in the center-of-mass motion, one can eliminate the cavity mode degree of freedom $\aop$ and write an effective master equation for the reduced density matrix of the sphere. This master equation leads to the following rate equation for the phonon number of the sphere center-of-mass degree of freedom:
\aleqn{
\frac{d \avg{\bdop \bop}}{dt} = -(A_{-}-A_+) \avg{\bdop \bop} + A_+
}
with the additional assumption that the bath attached to the cavity mode is at zero temperature. In this case the damping and driving terms of the above rate equation are given by 
\aleqn{
A_{\mp} &=  2g_l^2 \vert \alpha_1 \vert^2 \frac{\kappa_1}{\left(\Delta \pm \omega_1 \right)^2+\kappa_1^2},
}
with the shifted detuning $\Delta = \Delta'+2g_l^2\vert \alpha_1 \vert^2/\omega_1$. If at $t=0$ the initial number of phonons is $n_0$ (achieved \eg~via feedback cooling), the number of phonons at a later time $t_1$ is given by
\be 
\bar n = \frac{A_+}{A_{-}-A_+} + \pare{n_0 - \frac{A_+}{A_{-}-A_+}} e^{-(A_{-}-A_+) t_1}.
\ee 
An additional heating source  in the center-of-mass motion at a rate $\Gamma$ can be easily included by replacing $A_+$ with $A_+ + \Gamma$. Note that when the coupling to the cavity mode is turned on for the cooling process, the equilibrium position of the sphere along $x$ shifts to the mean field value $\beta_1 \neq 0$. One can expect that as the coupling is turned off slowly the cooled state of the particle will be adiabatically shifted to $\tilde{x} = 0$.

At $t=0$ the center-of-mass motion along the $x$-axis is thus prepared in a Gaussian state given by
\be \label{eq:SolStep1}
\begin{split}
v_x(t=0) &= \sigma_1^2 (2 \bar n +1), \\   
v_p(t=0) &= \frac{\hbar^2}{4 \sigma_1^2} (2 \bar n +1), \\ 
c(t=0) &= 0. \\ 
\end{split}
\ee 
This state has a purity $\mathcal{P} = 1/(2 \bar n+1)$ and an initial coherence length given by
$\xi(0) = \sqrt{8} \sigma_1/\sqrt{2 \bar n+1}$. 
Owing to the large mass, the coherence length at this step will be typically much smaller than the size of a single atom.  Our goal is to expand it to a value of the order of the radius of the sphere, that is, by many orders of magnitude. This is what we do in the next two steps of the protocol.

\subsection{Step 2: Boost}
As discussed in~\secref{Sec:Challenges}, in free space the coherence length of the state \eqref{eq:SolStep1} grows with a speed given by $\lim_{t \rightarrow \infty}\dot \xi(t) =2 (2 \bar n +1)^{-1/2} (\hbar \w_1/M)^{1/2}$. In order to increase this speed, in step 2 (boost) the center-of-mass evolves in an inverted harmonic potential of frequency $\w_2$~\cite{RomeroIsart2016}, given by the Hamiltonian $\Hop_2 = \pop^2/(2M)-M^2 \w_2^2 \xop^2/2$, from $t=0$ to $t=t_2$ such that $v_p$ is boosted by a factor of $\exp[2 \w_2 t_2]$. Assuming PLD in the LW limit, with a localization parameter $\Lambda_2$, the dynamics preserves the Gaussian character of the state at $t=0$ since the Hamiltonian is quadratic and the jump operators in the master equation are linear in $\xop$ and $\pop$. Thus the parameters characterizing the Gaussian state at $t=t_2$ can be obtained by analytically solving
\be \label{eq:Dynamics}
\begin{split}
\dot v_x (t) &=\frac{2}{M} c(t), \\
\dot v_p(t) &= 2 M \omega_2^2 c(t) + 2 \hbar^2 \Lambda_2 , \\
\dot c (t)&= M \omega_2^2 v_x(t)+ \frac{1}{M} v_p(t).
\end{split}
\ee
with the initial conditions given by \eqref{eq:SolStep1}. This leads to
{
\be 
\begin{split}
v_x(t_2) =&\frac{1}{2 }\spare{-\frac{v_p(0)}{M^2\omega_2^2}+v_x(0)-\frac{2\Lambda_2\hbar^2}{M^2\omega_2^2}t_2}\\
&+\spare{\frac{v_p(0)}{M^2\omega_2^2}+v_x(0)}\frac{\cosh\pare{2t_2\omega_2}}{2}\\
&+\spare{\frac{2c(0)}{M\omega_2}+\frac{\Lambda_2 \hbar^2}{M^2\omega_2^3}}\frac{\sinh\pare{2t_2\omega_2}}{2},
\end{split}
\ee
\be 
\begin{split}
v_p(t_2) =& \frac{1}{2}\spare{v_p(0)-v_x(0)M^2\omega_2^2+2\Lambda_2\hbar^2t_2} \\
&+\spare{v_p(0)+v_x(0)M^2\omega_2^2}\frac{\cosh\pare{2t_2\omega_2}}{2}\\
&+\spare{2c(0)M\omega_2+\frac{\Lambda_2\hbar^2}{\omega_2}}\frac{\sinh\pare{2t_2\omega_2}}{2},
\end{split}
\ee
\be 
\begin{split}
c (t_2)=& -\frac{\Lambda_2\hbar^2}{2 M \omega_2^2}\\
&+\spare{2c(0)+\frac{\Lambda_2\hbar^2}{M\omega_2^2}}\frac{\cosh\pare{2t_2\omega_2}}{2}\\
&+\spare{\frac{v_p(0)}{M\omega_2}+v_x(0)M\omega_2}\frac{\sinh\pare{2t_2\omega_2}}{2}.
\end{split}
\ee
}
Note the exponential dependence on $t_2 \w_2$ due to dynamics in an inverted potential.

\subsection{Step 3: Free}

After boosting the momentum fluctuations of the center of mass along the $x$-axis, the center-of-mass motion is let evolve freely from $t=t_2$ to $t=t_2+t_3$ in the presence of PLD with a localization parameter $\Lambda_3$. Denoting $T_m = \sum_{n=2}^m t_n$, at $t=T_3$ the time-evolved Gaussian state is given by solving \eqcite{eq:Dynamics} with $\w=0$, one obtains
\be
\begin{split}
v_x(T_3) =& v_x(t_2) + \frac{2c(t_2)}{M} t_3 + \frac{v_p(t_2)}{M^2}t_3^2 \\
&+ \frac{2 \Lambda_3 \hbar^2}{3 M^2} t_3^3, \\
v_p(T_3) =& v_p(t_2) + 2 \Lambda_3 \hbar^2 t_3, \\
c(T_3) =& c(t_2) + \frac{v_p(t_2)}{M} t_3 +\frac{\Lambda_3 \hbar^2 t_3^2}{M}.
\end{split}
\ee
Note that $(v_x)^{1/2} \sim t_3$ grows linearly in time until decoherence due to the $\Lambda_3$ term dominates, then $(v_x)^{1/2} \sim t_3^{3/2}$. In this regime the purity decays as $\mathcal{P} \sim t_3^{-2}$ causing the coherence length decay as $\xi \sim t_3^{-1/2}$. Note that so far no approximations have been made with regards to the treatment of the dynamics and thus the state of the center-of-mass can be fully determined. The difference in the size of the coherence length with or without the presence of gravitationally-induced decoherence is expected to be apparent at this point. For further details analyzing the advantage of using the inverted potential in step 2, see~\cite{RomeroIsart2016}.

\subsection{Step 4: Split}

In this step the spatially delocalized wavefunction is split by means of a double slit into a spatial superposition of two localized wavepackets. As discussed in \secref{Sec:Challenges}, it would be convenient to obtain a slit separation $d$ smaller than the diameter of the sphere. We propose to achieve this by performing a continuous time quantum measurement of $\xop^2$. The measurement outcome will determine the slit separation, and the strength of the measurement the width of the slits. The idea of using an $\xop^2$ measurement of a nanosphere to prepare a quantum superposition state was introduced in~\cite{ORI2011,ORI2011b}. In those cases the measurement was approximated to be instantaneous, here we address the more general situation considering that the measurement takes a finite time $t_4$. 

The measurement is modeled as follows. We consider the sphere to be quadratically coupled with a cavity mode that is resonantly driven with a strength $E_4$. The quadratic coupling strength is given by $g_q$, and in this step it is convenient to use the length unit $\sigma_4 = \xi(T_3) = \mathcal{P}_3 \sqrt{8 v_x(T_3)}$, namely we define $\tilde x=\xop/\sigma_4$. Here $\mathcal{P}_3$ is the purity of the state after free expansion \ie $\mathcal{P}(T_3)$. The Hamiltonian in the frame rotating at the cavity resonance frequency is given by
\aleqn{
\Hop_{4} = \frac{\pop^2}{2M} + \hbar g_q \tilde x^2 \adop \aop - i \hbar E_4 \left( \aop - \adop \right).  \label{eq:cavhamfull}
}
The decay rate of the cavity is given by $\kappa_4$. We assume that a homodyne measurement of the phase quadrature of the output cavity field is performed, such that the evolution of the density matrix $\hat \rho_T$ describing the center-of-mass motion and the cavity mode is given by the following stochastic master equation \cite{MilburnWisemanBook}
\be \label{eq:homdynesme}
\begin{split} 
\text{d} \hat \rho_T  = & -\frac{\im}{\hbar}\coms{\Hop_{4}}{\hat \rho_T}\text{d} t - \Lambda_4  \com{\xop}{\com{\xop}{\hat \rho_T}}\text{d} t \\
&+ 2 \kappa_4 \pare{ \aop\hat \rho_T \adop -\frac{1}{2} \coms{\adop \aop}{\hat \rho_T}_+} \text{d} t\\
& - \im \sqrt{2 \kappa_4}  \pare{ \aop \hat \rho_T - \hat \rho_T \adop - \avg{\aop - \adop} \hat \rho_T} \text{d} \mathcal{W},
\end{split} 
\ee 
where $\text{d} \mathcal{W}$ is the Wiener increment. Again, we assume PLD with a localization parameter $\Lambda_4$. After a displacement operation on the cavity field by an amount given by the classical field amplitude in the empty cavity, namely $\alpha_4 = -E_4/\kappa_4$, we can perform an adiabatic elimination (see \cite{Wiseman93,Corney98}) of the cavity mode when $g_q \abs{\alpha_4}/(2\kappa_4) \ll 1$.
The resulting equation for the reduced density matrix of the $x$-coordinate of the center-of-mass position of the sphere is given by
\be  \label{eq:SMEpos2}
\begin{split}
d \hat \rho =& -\frac{\im}{\hbar}\com{\frac{\pop^2}{2M} +   \frac{1}{2}M \w_g^2 \xop^2 }{\hat \rho} \text{d}t  \\
&- \Lambda_4  \com{\xop}{\com{\xop}{\hat \rho}}\text{d} t \\
&  -\lambda \com{\tilde x^2}{\com{\tilde x^2}{\hat \rho}} \text{d}t \\
& + \sqrt{2\lambda} \pare{\tilde x^2 \hat \rho+\hat \rho \tilde x^2-2\avg{\tilde x^2} \hat \rho} \text{d} \mathcal{W}.
\end{split}
\ee 
The effective measurement strength representing the continuous measurement of $\tilde x^2$ \cite{JacobsRev} is given by $\lambda \equiv g_q^2 |\alpha_4|^2/\kappa_4$. The quadratic coupling leads to an effective harmonic potential of frequency given by $  \w_g^2  \equiv 2\hbar   g_q \abs{\alpha_4}^2 /(M \sigma_4^2)$. 

To solve \eqnref{eq:SMEpos2} it is useful to write the density matrix in the position basis as
\be \label{eq:Aphi}
\bra{\tilde x} \hat \rho(t) \ket{\tilde x'} = A(\tilde x,\tilde x',t) \exp \spare{\im \phi(\tilde x,\tilde x',t)},
\ee 
where $A \in \mathbb{R}$, $\phi \in \mathbb{R}$, such that $A(\tilde x,\tilde x',t) = A(\tilde x',\tilde x,t)$, and $\phi(\tilde x,\tilde x',t) =- \phi(\tilde x',\tilde x,t)$. Note that since $\tr[\hat \rho^2] \leq 1$, then $\int \text{d} \tilde x \text{d} \tilde x' A^2(\tilde x, \tilde x',t) \leq 1$. By introducing \eqnref{eq:Aphi} in \eqnref{eq:SMEpos2}, one obtains 
\be  \label{eq:diffeqs1}
\begin{split}
\text{d} A  =& -\w_\sigma A \pare{\partial_{\tilde x}^2\phi - \partial_{\tilde x'}^2 \phi  }\text{d} t \\
& -2 \w_\sigma \spare{ \partial_{\tilde x}A \partial_{\tilde x}\phi -  \partial_{\tilde x'}A \partial_{\tilde x'}\phi     } \text{d} t \\
&-\Lambda_4 \sigma^2_4 (\tilde x-\tilde x')^2A \text{d} t\\
&-\lambda (\tilde x^2-\tilde x'^2)^2 A \text{d} t\\
&+ \sqrt{2\lambda} \pare{\tilde x^2+\tilde x'^2-2\avg{\tilde x^2} }A \text{d} \mathcal{W},
\end{split}
\ee 
and
\be  \label{eq:diffeqs2}
\begin{split}
\text{d}  \phi =& -\w_\sigma \spare{\pare{\partial_{\tilde x}\phi}^2-\pare{\partial_{\tilde x'}\phi}^2} \text{d} t \\
&+\frac{\w_\sigma}{A} \pare{\partial^2_{\tilde x}-\partial^2_{\tilde x'} }A \text{d} t \\
& -\frac{\omega^2_g}{4\w_\sigma} \pare{\tilde x^2 - \tilde x'^2} \text{d} t.
\end{split}
\ee 
where we have used $\w_\sigma \equiv \hbar/(2 M \sigma_4^2)$. At this step, one can perform the phase-amplitude separation approximation which corresponds to neglecting the second term in \eqnref{eq:diffeqs2}, thereby making the evolution of $\phi$ independent of $A$. This is justified since $\phi$ evolves much faster than $A$ as initially $\phi(\tilde x,\tilde x',T_3) =  \pare{\tilde x^2-\tilde x'^2} \Theta$ with $\Theta\equiv c(T_3)/(2\hbar) \gg 1$. Recall that to keep the purity constant, one requires $c^2(T_3)$ to be comparable to $v_x(T_3) v_p(T_3)$ which at $T_3$, after expanding the coherence length many orders of magnitude, is much larger than $\hbar^2/4$.  Using $\phi(\tilde x,\tilde x',T_3)$ as the initial condition, $\phi$ is given by $\phi(\tilde x,\tilde x',T_4)= F(t_4)(\tilde x^2-\tilde x'^2) $ with
\be  \label{eq:soldiffeqs3}
\begin{split}
F(t)= -\frac{\omega_g}{4\omega_\sigma}\frac{-4\Theta \omega_\sigma + \omega_g \tan{\omega_g t}}{\omega_g+4\Theta \omega_\sigma \tan{\omega_g t}}.
\end{split}
\ee 
Using this one can thus solve \eqnref{eq:diffeqs1} to obtain the time-evolved density matrix. By neglecting the PLD term and the contribution of phase gradients to the amplitude dynamics in \eqnref{eq:diffeqs1}, which lead to negligible contributions for the experimental parameters given in \secref{Sec:CaseStudy}, and enforcing that  $F(t_4) \approx 0$, which requires some fine tuning, the density matrix at $t_4$ is given by
\be
\rho(\tilde x,\tilde x',T_4) \approx M_{\mathcal{W}}(\tilde x) M_{\mathcal{W}}( \tilde x')  A(\tilde x,\tilde x',T_3),
\ee 
where
\be \label{eq:MOperator}
M_{\mathcal{W}}(\tilde x) = \pare{\frac{2}{\pi}}^{1/4}\exp \spare{ - \pare{\chi_4 \tilde x^2 - \frac{\mathcal{W}}{2 \sqrt{t_4}}}^2}.
\ee 
 Here $\mathcal{W}$ is a normally distributed random number with a variance $t_4$ and a zero-mean that parametrizes the outcome $p_L \equiv \mathcal{W}/(2 \sqrt{t_4})$ of the homodyne measurement. The measurement strength is given by $\chi_4 \equiv \sqrt{2\lambda t_4} = g_q \abs{\alpha_4} \sqrt{2 t_4/\kappa_4} \ll 2 \sqrt{t_4 \kappa_4}$, where the inequality is for the adiabatic elimination condition to hold. Note that the position probability distribution after the action of the measurement operator is given by two Gaussian peaks separated by
\aleqn{
d  =  2\sigma_4 \sqrt{\frac{p_L \chi_4 -\mathcal{P}_3^2}{\chi_4^2}} \approx 2 \sigma_4 \sqrt{\frac{p_L}{\chi_4}}, \label{eq:slitwidth}
}
and a width given by
\be \label{eq:peakwidth}
\begin{split} 
\sigma_{d}  &= \frac{\sigma_4}{\sqrt{8(p_L \chi_4-\mathcal{P}_3^2)}} \approx \frac{\sigma_4}{\sqrt{8p_L \chi_4}} = \frac{\sigma_4^2}{\sqrt{2} \chi_4 d}. 
\end{split}
\ee
We have assumed here that the measurement outcome is such that $p_L \chi_4 \gg \mathcal{P}_3^2$, which is the condition required to make the two peaks distinguishable. Thus the continuous time measurement of $\xop^2$ effectively acts as a double slit. We remark that the measurement operator~\eqnref{eq:MOperator} obtained within the phase-amplitude separation approximation has the same form as the one obtained assuming an instantaneous measurement~\cite{ORI2011}, however here we have derived the measurement strength as a function of the measurement time.

The probability distribution for the outcomes $P_o(p_L)$, when the measurement operator is applied on the input state $\hat{\rho}(T_3)$, is given by:
\aleqn{
P_o(p_L) &= \Tr \spare{\mcop_{\mathcal{W}} \hat \rho(T_3) \mcop_{\mathcal{W}}^{\dagger}} \nonumber \\
&= \pare{\frac{2}{\pi}}^{1/2}\int d \tilde{x} e^{-2\pare{\chi_4 \tilde{x}^2-p_L}^2}\langle \tilde{x} \vert \hat \rho(T_3) \vert \tilde{x} \rangle \label{eq:outprobdist}.
}
One aspect of realising such a double slit is that the slit separation \eqnref{eq:slitwidth} depends on the measurement outcome $p_L$ and will vary from run to run. As a result in our proposal we have to post-select a suitable slit-width that satisfies $\xi_G \ll d \lesssim \xi$ which will enable us to falsify gravitationally-induced decoherence. To this end it is advantageous also to have at hand the probability distribution for the slit separation $d$ for a given measurement strength given by
\aleqn{
P_s(d) = \frac{\chi_4 d}{2\sigma_4^2} P_o \pare{\frac{\chi_4 d^2}{4\sigma_4^2}+\frac{\mathcal{P}_3^2}{\chi_4}}. \label{eq:slitdist}
}
Using the above distribution we can now estimate the probability for the slit separation to fall in a given range of interest $d_{\mathrm{min}}<d<d_{\mathrm{max}}$. Choosing the minimum slit separation as $d_{\mathrm{min}}/\sigma_d = 5$ ensures that the two peaks in the post-measurement state are well resolved, while the condition $d \lesssim \xi$ naturally leads to the choice $d_{\mathrm{max}} = \xi(T_3)$. In the limit of large enough measurement strengths such that all the corresponding outcomes satisfy $p_L \chi \gg \mathcal{P}_3^2$, we get for this probability
\begin{align}
\int_{d_{\mathrm{min}}}^{d_\mathrm{max}} \text{d} l P_s(l) \approx \erf \, [\mathcal{P}_3] - \ \erf\,\left[\frac{2\mathcal{P}_3}{\sqrt{\chi_4}} \right],
\end{align}
which for large measurement strength $\chi_4 \gg 1$, tends to approach $\erf[1]\approx 0.8$.

At this step, the state after a double slit of separation $d$ has been prepared. As discussed in \secref{Sec:Challenges}, the generation of fringes in free expansion would require an unfeasibly long time. In the next two steps, we propose a method to overcome this challenge.

\subsection{Step 5: Rotation}

In the next two steps, step 5 (rotation), and step 6 (inflation), we will ensure that the fringes separate exponentially in time instead of linearly as implied by dynamics in free space~\cite{RomeroIsart2016}. To achieve this, first, in step 5 (rotation), we let the system evolve with the Hamiltonian $\Hop_5 = \pop^2/(2M) + M \omega_5^2 \xop^2/2$ during a time $t_5$, such that $\xop$ and $\pop$ evolve in the Heisenberg picture as
\be \label{eq:Rotation}
\begin{split}
\xop(T_5)&=\xop (T_4) \cos(\omega_5 t_5)+\frac{\pop(T_4)}{M \omega_5}\sin(\omega_5 t_5),\\
\pop(T_5)&=\pop(T_4) \cos(\omega_5 t_5) -M\omega_5 \xop(T_4) \sin(\omega _5 t_5).
\end{split}
\ee
Here, for simplicity, we have ignored decoherence for the moment. In step 6 (inflation), the system evolves with the Hamiltonian $\Hop_6 = \pop^2/(2M) - M \omega_6^2 \xop^2/2$ for a time $t_6$,   such that $\xop$ and $\pop$ evolve in the Heisenberg picture as
\be 
\begin{split}
\xop(T_6)&=\xop (T_5) \cosh(\omega_6 t_6)+\frac{\pop(T_5)}{M \omega_6}\sinh(\omega_6 t_6),\\
\pop(T_6)&=\pop(T_5) \cosh(\omega_6 t_6) + M\omega_6 \xop(T_5) \sinh(\omega_6 t_6).
\end{split}
\ee
Note that for $t_6 \gg \w_6^{-1}$, then
\be 
\xop(T_6) \approx \frac{e^{\omega_6 t_6 }}{2} \spare{\xop (T_5) +\frac{\pop(T_5)}{M \omega_6}}.
\ee
Using \eqcite{eq:Rotation}, and fine tuning $t_5$ such that 
\be \label{eq:tof_condition}
\cos ( \w_5 t_5) = \frac{\w_5}{\w_6} \sin (\w_5 t_5),
\ee
then
\be
\xop(T_6) \approx  e^{\omega_6 t_6 } \pop(T_4) \frac{\sin(\w_5 t_5)}{2 M \w_5} \spare{1+\pare{\frac{\w_5}{\w_6}}^2}.
\ee
Thus $\xop(T_6)$ is proportional to $\pop(T_4)$ with a coefficient that grows exponentially with $\w_6 t_6$.  Therefore one can perform an exponentially fast time-of-flight measurement of the momentum distribution at $t=T_4$ by measuring the position distribution at $t=T_6$. Note that for $\w_5=\w_6$ the condition \eqcite{eq:tof_condition} is satisfied with $\w_5 t_5 = \pi/4$. We remark that should one be interested in measuring the probability distribution of another quadrature at $t=T_4$, one could adjust $t_5$ accordingly.

Motivated by the discussion above, let us calculate the state obtained after rotation for a time $t_5$. Since after step 4 (split), the state is non-Gaussian, its evolution has to be calculated more carefully. In this context, it is convenient to use the Wigner function
\be
 W(x,p,t)= \int_{- \infty}^\infty \text{d}y \frac{e^{- \im p y/\hbar}}{2 \pi \hbar}  \bra{x +\frac{y}{2}} \hat \rho(t) \ket{x-\frac{y}{2}}.
\ee
In the presence of PLD in the LW limit, with localization parameter $\Lambda_5$, and the Hamiltonian $\Hop_5$, the evolution of the Wigner function is given by
\be 
\begin{split}
\frac{\partial W}{\partial t}=-\frac{p}{M}\frac{\partial W}{\partial x}+M \omega_5^2 x\frac{\partial W}{\partial p}+\hbar^2\Lambda_5 \frac{\partial^2W}{\partial p^2}.
\end{split}
\ee
To solve this eqution it is useful to define the Fourier transform
\be 
\bar{W} (k_x,k_p,t) =\int_{-\infty}^{\infty}W(x,p,t)e^{-i k_x x}e^{-i k_p p}\text{d}x \text{d}p,
\ee
such that the dynamical equation reads
\be 
\begin{split}
\frac{\partial \bar{W}}{\partial t}=\frac{k_x}{M}\frac{\partial \bar{W}}{\partial k_p}-M \omega_5^2 k_p\frac{\partial \bar{W}}{\partial k_x}-\hbar^2\Lambda_5 k_p^2\bar{W}.
\label{recpW}
\end{split}
\ee
Let us now make the ansatz 
\be 
\begin{split}
 \bar W_{\Lambda\neq0}(k_x,k_p,t)=  \bar {W}_{\Lambda=0} (k_x,k_p,t) \bar G ( k_x,k_p,t),
\label{antz}
\end{split}
\ee
where $ \bar W_{\Lambda = 0}(k_x,k_p,t)$ ($ \bar W_{\Lambda\neq0}(k_x,k_p,t)$) is the solution of \eqcite{recpW} with $\Lambda=0$ ($\Lambda \neq 0$). 
Plugging \eqnref{antz} into \eqnref{recpW} leads to the following equation for the convolution function $\bar{G}$
\be 
\begin{split}
\frac{\partial \bar{G}}{\partial t}=\frac{k_x}{M}\frac{\partial \bar{G}}{\partial k_p}-M \omega_5^2 k_p\frac{\partial \bar{G}}{\partial k_x}-\hbar^2\Lambda_5 k_p^2\bar{G}.
\label{recpG}
\end{split}
\ee
This equation together with the initial condition $\bar G(k_x,k_y,0) = 1$ has the following solution
\be 
\begin{split}
\bar{G}=&\exp \spare{ \frac{\hbar^2\Lambda_5}{ 4 M^2\omega_5^3} g(k_x,k_p,t)},
\label{solrecpG}
\end{split}
\ee
where
\be 
\begin{split} 
 g(k_x,k_p,t) & \equiv\pare{k_x^2+k_p^2 M^2 \omega_5^2} \times \\
& \left\{ -2 t \omega_5 -\sin \spare{2\arctan\pare{\frac{M \omega_5 k_p}{k_x}}} \right.\\
&\left. +\sin \spare{2t  \omega_5 + 2\arctan \pare{\frac{M \omega_5 k_p}{k_x}}} \right\}.
\end{split}
\ee 
This gives the exact solution of the evolved Wigner function including PLD in the LW. Note that for a quadratic Hamiltonian the evolution of the Wigner function in the absence of decoherence is given by $W_{\Lambda =0} (x,p,t)=W_{\Lambda =0} (x(t),p(t),0)$, where $x(t)$ and $p(t)$ are the classical solutions given by the Hamilton equations.

In order to quantify the effect of decoherence during the rotation, it is convenient to define dimensionless variables: $\bar k_x\equiv \sigma_x k_x$, $\bar k_p\equiv \sigma_p k_p$ and  $\bar t\equiv \omega_5 t_5$, for some given length and momentum units $\sigma_x$ and $\sigma_p$. Then,
\be 
\begin{split}
\bar{G}=&\exp \spare{ -(A_1\bar k_x^2+A_2 \bar k_p^2 )B(\bar k_x,\bar k_p,\bar t)}
\end{split}
\ee
with
\be 
\begin{split}
A_1\equiv& \frac{\hbar^2\Lambda_5}{ 4 M^2\omega_5^3\sigma_x^2},\\
A_2\equiv& \frac{\hbar^2  \Lambda_5}{ 4 \omega_5\sigma_p^2},\\
B\equiv& 2\bar t +\sin \spare{2\arctan \pare{\frac{M \omega_5\sigma_x }{\sigma_p}\frac{\bar k_p}{\bar k_x}}}\\
&-\sin \spare{2\bar t +2 \arctan \pare{\frac{M \omega_5\sigma_x }{\sigma_p}\frac{\bar k_p}{\bar k_x} }}. \\
\end{split}
\ee
With the experimental parameters used in the case study \secref{Sec:CaseStudy}, and with $\sigma_x = d$ and $\sigma_p = \hbar/\sigma_d$ (recall \eqcite{eq:slitwidth} and \eqcite{eq:peakwidth}),  one can show that typically $A_1, A_2 \ll 1$ and thus decoherence in this step can be neglected.

\subsection{Step 6: Inflation}

In step 6, the center of mass evolves in an inverted potential of the form $\Hop_6 = \pop^2/(2M) - M \w_6^2 \xop^2/2$ and PLD with localization parameter $\Lambda_6$ from $t=T_5$ to $t=T_6$. The state can be calculated using the Wigner function, as discussed in step 5 replacing $\omega_5 \rightarrow i \omega_6$ and $\Lambda_5 \rightarrow \Lambda_6$ . 

After the evolution, we expect to observe fringes in the position probability distribution, namely in 
$ P \pare{x}\equiv\int_{-\infty}^{\infty}W \pare{x,p,T_6}dp$. Using the results of the previous section, one can show 
\be \label{eq:Fringes}
P_{\Lambda\neq 0}\pare{x}=\int_{-\infty}^{\infty}P_{\Lambda= 0}\pare{x+y}\frac{\exp\pare{-y^2/\sigma_{\Lambda}^2}}{\sqrt{\pi}\sigma_{\Lambda}}dy,
\ee
where under the action of $\Hop_6$
\be \label{eq:blurring}
\begin{split}
\sigma_{\Lambda}= \sqrt{\frac{\hbar^2\Lambda_6}{ M^2\omega_6^3}\spare{\sinh\pare{2t_6\omega_6}-2 t_6\omega_6}}
\end{split}
\ee
is the blurring distance due to decoherence. 
One can show that the fringe separation grows exponentially in time approximately as $x_f \approx \exp[t_6 \w_6]2 \pi \hbar/(Md \w_6)$. It will thus be important that decoherence during the inflation step is low enough such that $\sigma_\Lambda/x_f \ll 1$ at $T_6$, and hence the interference pattern is not washed out.

Note that the position probability distribution of the center-of-mass after the step 6 (inflation) can be calculated analytically (even though the final expression is very lengthy) taking into account all the experimental parameters and PLD in the LW limit. This is used in \secref{Sec:CaseStudy} to obtain the interference pattern for a specific experimental implementation of the protocol.

\subsection{Step 7: Measure}

The final step (measure) consists in performing a continuous time measurement of $\xop^2$ in order to unveil the fringes \eqcite{eq:Fringes}. Note that since the interference pattern has to be symmetric, it is more convenient to directly measure $\xop^2$, as is done in step 4 (split). Hence, in this step the Hamiltonian is given by
\aleqn{
\Hop_{7} = \frac{\pop^2}{2M} + \hbar g_q \tilde x^2 \adop \aop - i \hbar E_7 \left( \aop - \adop \right),
}
where the convenient length scale unit is now given by $\sigma_7 =\avg{\xop^2}^{1/2}=[ \int_{-\infty}^\infty x^2 W(x,p,T_6) \text{d}x \text{d}p]^{1/2}$. Note that the cavity parameters in this step are given by: the resonant driving strength $E_7$, the decay rate of the cavity given by $\kappa_7$, and the classical field amplitude in the empty cavity $\alpha_7= - E_7/\kappa_7$. A homodyne measurement of the phase quadrature of the cavity field is performed assuming, again, the adiabatic approximation, namely $g_q^2 \abs{\alpha_7}^2/(2 \kappa_7) \ll 1$. Further, invoking the phase-amplitude separation approximation, which is valid again due to the large phase accumulated during step 5 (inflation), the measurement strength is given by $\chi_7= g_q \abs{\alpha_7} \sqrt{2 t_7/\kappa_7} \ll 2 \sqrt{t_7 \kappa_7}$. The position resolution that can be achieved in the measurement is bounded from below by
\be 
\delta x \ge \frac{\sigma_7}{\sqrt{2 \chi_7}} \gg \frac{\sigma_7}{2 (t_7 \kappa_7)^{1/4}}.
\ee 
This position resolution ignores measurement back-action which is justified in the parameter regime we are interested in. In order to resolve the fringes we require $x_f \gg \delta x$.

\section{quantum micromechanical interferometer: implementation}\label{Sec:QIImplementation}

\begin{figure*}[t]
\centering
\includegraphics[width=2 \columnwidth]{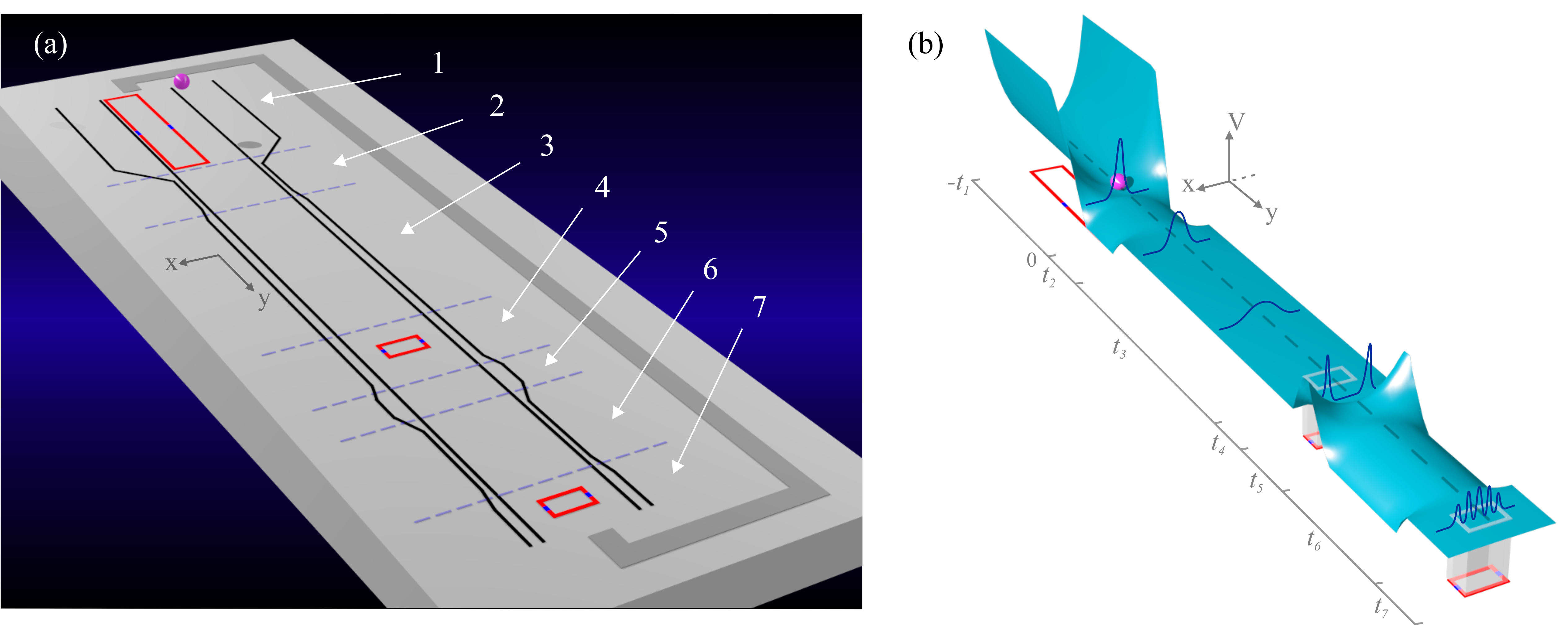}
\caption{ (a) Sketch of the superconducting chip implementation of the quantum micromechanical interferometer protocol. Superconducting wires are shown in black and the different stages are separated by dashed lines. The SQUIDs are shown in red.  (b) Illustration of the magnetic potential $V(x,y)$ in each of the steps. The position probability distribution in the $x$-axis is illustrated (dark blue) at the different stages of the protocol. Notice that both figures are not scaled and have only illustrative purposes.}
\label{Fig3}
\end{figure*}

In this section, we propose and analyze a physical set-up for the implementation of the quantum micromechanical interferometer protocol discussed in \secref{Sec:QIProtocol}, see \figref{Fig3}. We consider a superconducting sphere that is magnetically levitated on top of a superconducting chip, at some distance $z=z_t$, and skates along the $y$-axis.  The main ingredients of the implementation are the following:
\begin{itemize}
    \item The different steps of the protocol are implemented by letting the sphere skate along the $y$-axis in a static potential landscape as sketched in \figref{Fig3}b. The potentials along the $x$-axis, $V(\xop)$, act during the time required for the sphere to skate through the particular region in the potential landscape. This is advantageous since the potential is static and does not need to be switched-on and off, which could easily cause additional decoherence.
    
    \item The potential landscape is generated by persistent superconducting currents that give rise to a static magnetic field potential that interacts with a superconducting microsphere in the Meissner state~\cite{RomeroIsart2012}. Being persistent, the intensity of the currents does not fluctuate and hence the magnetic potentials are stable, minimising decoherence arising from trap fluctuations.
    
    \item The cavities required for cooling, splitting, and measuring the position of the state are implemented with flux-dependent microwave quantum cavities containing a SQUID. This leads to a quantum magnetomechanical coupling~\cite{RomeroIsart2012,Cirio2012,Via2015} between the sphere and the cavities. These quantum circuits are positioned on a chip at $z=0$, where the persistent currents are located as well. The sphere skates along the $y$-axis, levitated at some distance $z=z_t$. The SQUIDs are positioned in the appropriate region of the magnetic landscape such that when the particle skates over them, the desired flux-dependent couplings to the particle position (or squared position) are implemented. Note that the driving fields of the cavities can be switched on throughout since the interaction of the sphere with the cavity will only take place while the sphere is skating near them.
\end{itemize}
In order to control the speed at which the sphere skates along the $y$-axis, the chip could be conveniently inclined. A superconducting sphere with optimal properties could be loaded on a superconducting chip using a magnetic conveyor as has been experimentally realized with ultracold atoms~\cite{Minniberger2014}. Being an on-chip implementation, one could imagine having a system in which the same sphere can be reused for each run of the experiment. At the end of each run the sphere could be brought back to step 1 with an on-chip magnetic conveyor. This would prevent the run-to-run variations that would necessitate an average over mass and magnetic properties of the different spheres used to obtain the interference pattern. 

In the following, we discuss how the magnetic landscape in \figref{Fig3}b can be implemented with persistent currents. Subsequently we discuss how steps 1, 4, and 7 can be performed using a quantum magnetomechanical coupling to a microwave resonator. Then we consider decoherence due to magnetic field fluctuations arising from the surface, a source of decoherence which is introduced by choosing this particular physical implementation.

\subsection{On-chip magnetic landscape}

\begin{figure*}[t]
\begin{center}
\includegraphics[width=2\columnwidth]{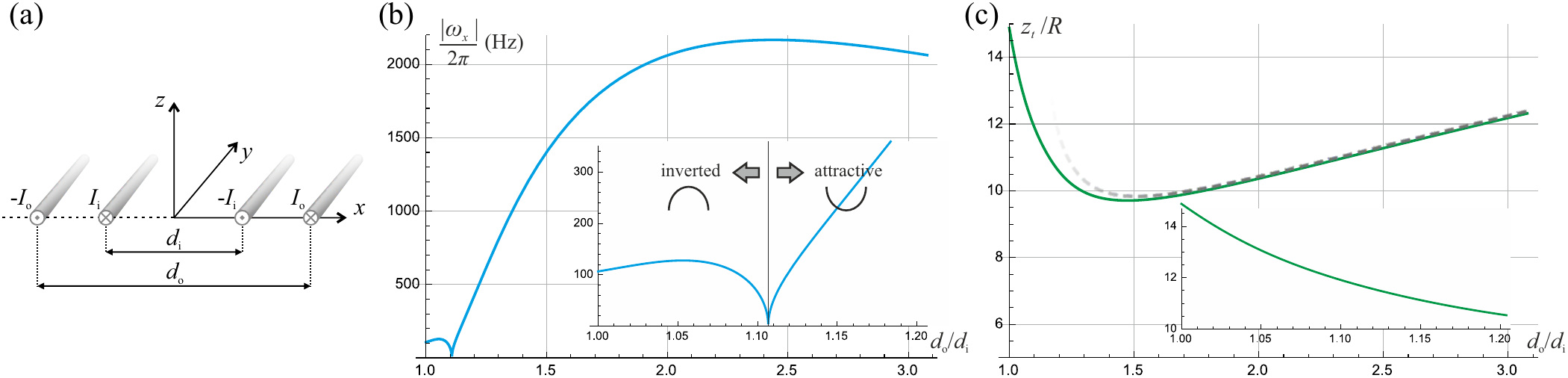}
\caption{(a) Sketch of the wires forming the building blocks of the magnetic skatepark. (b) Plot of the frequency of the harmonic/inverted potential in $x$ as a function of the ratio of distances between wires $d_o/d_i$ (fixing $d_i=13 \, R$) and currents specified in Appendix. The inset shows in detail the change in the potential from inverted to attractive. (c) Plot of the corresponding trapping heights $z_t/R$ (green line). The point of zero field, $z_{B=0}$, given by \eqnref{0fpoint}, is indicated with a dashed gray line.}
\label{Fig4}
\end{center}
\end{figure*}

 For a perfect superconducting sphere of volume $V$ in an applied magnetic field $\BB$, the Hamiltonian describing the interaction between the magnetic field and the induced dipole~\cite{jackson} of the sphere, whose center of mass is at position $\rr=(\xop,y,z)$ in a gravitational field along the $z$-axis, can be approximated by by~\cite{RomeroIsart2012}
\be \label{eq:MagPot}
\hat V(\rr)= \chi | \BB(\rr)|^2 +M g z,
\ee
where $\chi \equiv 3V/4 \mu_0 $,  $\mu_0$ is the vacuum magnetic permeability, and $g$ is the effective gravitational acceleration along the $z$-axis taking into account the small inclination of the chip. This expression assumes that the superconducting sphere is in the Meissner state, namely that $\BB =0$ in the interior of the sphere. {Such assumption requires the radius of the sphere to be much larger than the fundamental length scales in superconductivity, namely the London's penetration depth and the superconducting coherence length~\cite{tinkham}}. 

The magnetic field used to engineer the potential required in each of the steps of the protocol will be created by 4 parallel wires~\cite{revmodern,Thywissen}. Let us start by considering 4 ideal infinite wires (neglecting their thicknesses) placed along the $y$-direction at $z=0$. The two interior wires, separated by a distance $d_i$, carry a current of strength  $I_i$ in opposite directions. The two outer wires, separated by a distance $d_o$, carry a current $I_o$, as sketched in \figref{Fig4}a. These wires create a magnetic field, $\BB(\rr)$, that does not depend on $y$. In absence of gravity, analytical expressions of the trapping position can be derived \cite{revmodern,Thywissen}. 
Considering gravity, we numerically study the properties of the traps. In particular, we study the normalized potential seen by the sphere as a function of its coordinates $\hat V(\rr)/E_0=| \BB (\rr)|^2/B_0^2+\zeta z/R$,
where $B_0=\mu_0 2 I_i/(\pi d_i)$ is the field created by the two inner wires at the origin of coordinates, $E_0 \equiv \chi B_0^2$, and $\zeta \equiv 4 \mu_0 \rho g R/(3 B_0^2)$. Here
$\rho$ is the mass density of the sphere. The properties of the magnetic potential can be modified by tuning the distances and intensities of the wires. For the protocol discussed in \secref{Sec:QIProtocol}, the following potential distributions are needed:
\begin{itemize}
\item {Strong attractive potential (step 1)}: strong confinement in the $x$-direction, while the particle is trapped at $x_p=0$ and $z_p=z_t$.
\item {Inverted potential (steps 2, 6)}: inverted potential in the $x$ direction while the particle is levitating at $z_p \approx z_t$.
\item {Free propagation potential (steps 3, 4, 7)}: flat potential in the $x$-direction while the particle is levitating at $z_p \approx z_t$.
\item {Weak attractive potential (step 5)}: weak confinement in the $x$-direction, with a frequency value similar to that for the inverted potential while the particle is trapped at $x_p=0$ and $z_p \approx z_t$.
\end{itemize}
All of the different potentials required above can be obtained by fixing the current intensities through the wires and the distance between the interior wires, and only tuning the distance between the outer ones. In \figref{Fig4}b we plot the absolute value of the trapping frequency in $x$ as a function of the distance between the outer wires (for the particular distances and currents used in the case study \secref{Sec:CaseStudy}). In  \figref{Fig4}c we plot the corresponding trapping heights $z_t$. As can be seen from these plots, the magnetic potential can be  modified by only tuning the outer distance between wires, allowing to create the different distributions required for the protocol. The strong attractive potential is generated placing the outer wires at a distance $d_o \simeq 2.3 \,d_i$. This generates a quadrupole-like field distribution, and the trapping takes place slightly below the zero field point, $z_{B=0}$, given by~\cite{revmodern}
\begin{align}
z_{B=0}=\frac{d_i}{2}\sqrt{\frac{d_o/d_i-I_o/I_i}{I_o/I_i-d_i/d_o}}, \label{0fpoint}
\end{align}
A tight trap both in $x$ and $z$-directions is then created.  On the contrary, the inverted potential is obtained by placing the outer wires closer to the inner ones. In this case, as $d_{o}\rightarrow d_{i}$, the potential is the same as that created by a single pair of wires with current ($I_{i}-I_{o}$), giving rise to an inverted potential in $x$ and a stable trapping point in $z$ due to the gravitational force. Between these two cases, one can find a $d_{o}$ value for which the potential in the $x$ direction at $x=0$ changes from being inverted to attractive, see \figref{Fig4}b. This allows us to design a flat potential permitting free expansion of the state along the $x$-axis. A possible set of experimental numbers for implementing the protocol are discussed in \secref{Sec:CaseStudy}.

These results have been obtained assuming that the wires are infinitely long. However, we have performed calculations considering the complete protocol with the different finite regions as sketched in \figref{Fig3}a. Both trapping heights and potential profiles agree well with the results for infinite guides and transitions from one region to the other can be made adiabatic by controlling the lengths of the connections between the different regions. However, taking into account these smooth transitions in the protocol, and choosing the experimental parameters accordingly in the case study discussed in \secref{Sec:CaseStudy} lies beyond the scope of this manuscript.

\subsection{Cavity Quantum magnetomechanics}

In steps 1, 4, and 7 a coupling to a cavity mode is required. This is achieved by a magnetomechanical coupling between the center-of-mass of the superconducting microsphere and a microwave resonator with a SQUID~\cite{RomeroIsart2012,Cirio2012,Via2015}. The frequency of the cavity mode is flux-dependent due to the presence of the SQUID~\cite{Koch2007}, and it is given by
$ \Hop = \hbar \omega_c (\rr) \adop \aop$, where $\adop$ ($\aop$) is the creation (annihilation) cavity mode operator,
$\hbar \omega_c (\rr)  = [8 E_J(\rr) E_C]^{1/2}- E_C$, and
$
E_J(\rr) = 2 E_{J_1} \cos \spare { \pi \Phi (\rr)/\Phi_0 }.
$
Here $\Phi_0$ is the flux quantum, $\rr$ is the position of the superconducting sphere,  $E_C$ is the charging energy of a single electron stored in the capacitance,  and $E_{J_1}$ the energy associated with an electron tunneling across one of the two identical junctions. The regime $E_J / E_C \gg 1$ is assumed so that the resonator is linear. Note that the magnetomechanical coupling arises because the flux threading the pick-up coil $\Phi(\rr)$ depends on $\rr$. As before we denote the position of the sphere as $\rr=(\xop,y,z)$ with $\xop$ corresponding to the quantum mechanical position operator of the center-of-mass motion along the $x$-axis. Expanding the Hamiltonian up to second order in $\xop$, one has
\begin{align} \label{eq:QMMH}
\Hop= \hbar \omega_c  \adop \aop + \hbar g_{l} \tilde{x} \adop \aop + \hbar g_{q} \tilde{x}^2 \adop \aop,
\end{align}
where $\w_c \equiv \w_c(0,y,z)$, and $\tilde{x} = \xop/\sigma$ is the position scaled by a length scale which we choose in each step of the protocol and is related to $(\tr[\xop^2 \hat \rho])^{1/2}$. The linear and quadratic magnetomechanical couplings are then given by 
\be
\begin{split} 
g_l =&  - \frac{\pi \omega_0 s_0  \text{sign}[c_0]}{\sqrt{2 c_0}} \eta_l,\\
g_q =& -\frac{\omega_0 \text{sign}[c_0] }{2}   \spare{ \frac{\pi^2 \left(1+c_0^2\right)}{( 2  c_0  )^{3/2}} \eta_l^2 + \frac{\pi s_0 }{\sqrt{2 c_0} } \eta_q },
\end{split} 
\ee 
where we have defined the frequency $\hbar \w_0 \equiv (8 E_{J_1} E_c)^{1/2} $, $s_0 \equiv \sin[ \pi \Phi(0)/\Phi_0]$, $c_0 \equiv \cos[ \pi \Phi(0)/\Phi_0]$, $\Phi(0) = \Phi(0,y,z)$, and the magnetomechanical dimensionless parameters
\be
\begin{split} 
\eta_l \equiv & \frac{\sigma}{\Phi_0} \left. \fpd{\Phi(x,y,z)}{x} \right|_{x=0}, \\
\eta_q \equiv &\frac{\sigma^2}{\Phi_0} \left. \spd{\Phi(x,y,z)}{x} \right|_{x=0}.
\end{split} 
\ee 
Note that to have a quadratic coupling  the pick-up coil is placed such that $\eta_l=0$ and $s_0 \neq 0$. Whenever $\eta_l \neq 0$ and $s_0 \neq 0$, the linear coupling is much larger than the quadratic since $g_l/g_q \sim \eta_l/(\eta_l^2+\eta_q)$  and $1 \gg \eta_l \gg \eta_q$ for the Taylor expansion to be valid. 
Finally, the microwave photons in the cavity decay with a rate $\kappa \equiv \w_c/Q$, where $Q$ is the cavity quality factor. The dissipative term in the master equation describing this decay is given by
\be \label{dissterm}
\mathcal{L}[\hat \rho]= 2 \kappa \pare{\aop \hat \rho \adop - \frac{1}{2} \spare{\adop \aop, \hat \rho}_+}.
\ee 
Thus the Hamiltonian \eqref{eq:QMMH} and the dissipative term \eqref{dissterm} are of the same form as those used in \secref{Sec:QIProtocol} to discuss steps 1, 4, and 7. As was remarked in \cite{ORI2011}, the non-linear cavity cooperativity due to the quadratic coupling is greatly enhanced due to the large scale given by $(\tr[\xop^2 \hat \rho])^{1/2}$, many orders of magnitude larger than the typical zero point motion in quantum nanomechanical systems where non-linear couplings are typically very weak~\cite{revopt}. Note, however, that there has been recent experimental progress for position-squared measurements of mechanical systems~\cite{Doolin2014,Praiso2015,Brawley2016}. We remark that the SQUIDs are assumed noiseless. The effect of flux noise in the SQUIDs, in particular in the double slit in stage 4, will be addressed elsewhere~\cite{VenkateshPrep}.

Let us now discuss the appropriate location for the pick-up coils on the chip such that the steps 1,4, and 7 of the protocol are passively implemented in the magnetic skatepark. We choose to place the pick-up coils on the chip \emph{i.e.} in the $z=0$ plane. We consider rectangularly shaped coils with lengths along the $y$ and $x$-axis given by $l_y$ and $l_x$ respectively. The magnetic flux threading the pick up coil $\Phi$ has a contribution given by the field created by the induced currents of the superconducting sphere and a contribution given by the field created by the wires. With the classical position of the sphere given by $\rr=(0,0,z_t)$ and the center of the coil at the origin $\rr_c=(0,0,0)$, the first derivative of the flux respect to the $x$-position of the sphere is zero due to the symmetry of the coil and the magnetic field. Thus, to obtain a non-zero linear coupling, $\eta_l$, the coil has to be shifted from the center in the $x$ direction. Restricting to the available space between the inner wires forming the magnetic guide ($d_{\rm i}/R=13$ as used in the case study \secref{Sec:CaseStudy}), the linear coupling $\eta_l$ can be easily optimized over $x_c$ and $l_x$ resulting in  a maximum for 
$x_c/R=3$ and $l_x/R=6$. On increasing the length $l_y$ of the coil, the coupling increases  until  $l_y/R \simeq 40$ and saturates thereafter. Note that the total flux that crosses the coil, which determines the constant flux coefficient $s_0$ entering into the quantum magnetomechanical couplings, depends on its geometrical properties.  Regarding the quadratic coupling, $\eta_q$, we consider a centered coil in $x$, which ensures that the first derivative of the flux with respect to the $x$-position of the sphere is zero. Also in this case, both the magnetomechanical coefficient and the constant flux coefficient depend on the dimensions of the coil. The particular experimental numbers used in the case study \secref{Sec:CaseStudy} are given in the Appendix. We emphasize that the couplings are switched-on and off dynamically by letting the sphere skate over the area where the pick-up coils are located without changing the intensity of the driving fields, something that can be advantageous to minimize decoherence and noise.

\subsection{Magnetic field fluctuations}

In the on-chip implementation of the quantum micromechanical interferometer protocol, potential surface-induced sources of decoherence could play a role. In particular,
as the microsphere couples to the fluctuations of the electromagnetic (EM) field, both quantum and thermal, we expect its motion to decohere as a consequence. Such decoherence can be studied using the quantum Brownian motion (QBM) model~\cite{Breuer2002, Schlosshauer2007} with the center-of-mass motion as the system, coupled linearly to the bath of EM field modes. In this section, we consider the decoherence induced by the interaction of a diamagnetic particle with the fluctuations of the electromagnetic field near a surface, which could potentially change the density of bath modes to a large extent. Looking at the interaction between a diamagnetic particle and the field as in \eqnref{eq:MagPot}, in the presence of an externally applied classical magnetic field $\BB(\rr)$ with fluctuations $\hat{\BB}_f(\rr)$, one can write the interaction Hamiltonian up to bilinear terms as
\be \label{HintB}
\Hop_I \approx \chi\spare{ \BB\pare{\rr_0}\cdot\partial_x\hat{\BB}_f\pare{\rr_0} + \hat{\BB}_f\pare{\rr_0}\cdot\partial_x\BB\pare{\rr_0} }\hat{x},
\ee 
where $\rr_0$ is the classical trajectory of the particle, and $\hat{x}$ stands for the quantum fluctuations of the center-of-mass position along the $x$-axis. We have ignored here the contribution from the interaction terms of third and higher order in fluctuations of the magnetic field and  of the center-of-mass position.  One can then express the magnetic field fluctuations in the presence of the surface as \cite{SYB1}
\aleqn{\label{Bfluc}\hat{\BB}_f(\rr) =& \int_0^\infty \frac{\mathrm{d}\omega}{i\omega}\sum_{\lambda = e,m}\int \mathrm{d}^3 \rr' \vec{\bf \nabla}\times\bar{\bar {G}}_\lambda({\bf r},{\bf r}'\omega)\cdot\hat{\bf f}_\lambda ({\bf r}',\omega)\nonumber\\&+\hc,}
where $[\vec{\bf \nabla}\times\bar{\bar {G}}_\lambda({\bf r},{\bf r}'\omega)]_{il} = \epsilon_{ijk}\partial_j [\bar{\bar{G}}_\lambda(\rr,\rr',\omega)]_{kl}$. The operators $\hat{\ff}^{\dagger}_\lambda (\rr,\omega)$ and $\hat{\ff}_\lambda (\rr,\omega)$ refer to the bosonic creation and annihilation operators for the medium-assisted field that follow canonical commutation relations, and the coefficients $\bbar{G}_\lambda\pare{\rr_1,\rr_2,\omega}$  are defined as 
\aleqn{\bbar{G}_e \pare{\rr,\rr',\omega}= i\frac{\omega^2}{c^2} \sqrt{\frac{\hbar}{\pi\epsilon_0}\Im[\epsilon \pare{\rr',\omega}]} \bbar{G}\pare{\rr,\rr',\omega}\\
\bbar{G}_m \pare{\rr,\rr',\omega}= i\frac{\omega^2}{c^2} \sqrt{\frac{\hbar}{\pi\epsilon_0}\frac{\Im [\mu \pare{\rr', \omega}]}{\abs{\mu\pare{\rr',\omega}}^2}} \bbar{G}\pare{\rr,\rr',\omega},}
with $\epsilon(\rr,\omega)$ and $\mu(\rr,\omega)$ as the space-dependent  permittivity and permeability, and $\bbar{G}\pare{\rr_1,\rr_2,\omega}$ as the Green's tensor for a particle near a planar semi-infinite surface \cite{SYB1}. 
Rewriting the interaction Hamiltonian \eqref{HintB} as $H_I \equiv \hat{\mathcal{C}}\hat{x}$, where one can identify the operator   $\hat{\mathcal{C}}=\BB\pare{\rr_0}\cdot\partial_x\hat{\BB}_f\pare{\rr_0} + \hat{\BB}_f\pare{\rr_0}\cdot\partial_x\BB\pare{\rr_0} $ as the bath operator pertaining to the QBM model, one can readily see that the interaction will yield a position localization decoherence as the bath monitors the system's position coordinate via the linear coupling. Assuming that the particle is in a harmonic trap of frequency $\omega_t$, the decoherence resulting from the quantum Brownian motion can be written as~\cite{Breuer2002, Schlosshauer2007}
\aleqn{\Lambda_{B}  &= \frac{1}{2\hbar^2}\int_0^\infty \mathrm{d}\tau \mathcal{N}(\tau)\cos(\omega_t\tau), }
 where $\mathcal{N}(\tau)\equiv\avg{[\hat{\mathcal{C}},\hat{\mathcal{C}}(-\tau)]}$  stands for the noise correlation function of the thermal bath at temperature $T_e$. Expressing the magnetic field fluctuations in terms of the medium-assisted field operators and simplifying further, we obtain the decoherence to be
 \aleqn{\Lambda_B=  \frac{\pi }{2 \hbar}\coth\pare{\frac{\hbar \omega_{t}}{2K_b T}}J(\omega_{t}),}
  where the effective spectral density of the vacuum modes  $J\pare{\omega}$ in the presence of a surface is given as \cite{DIP}
\aleqn{J(\omega) =& \frac{\mu_0\chi^2}{\pi}\left[\partial_x\BB\cdot\bbar{\ca{G}}\pare{\rr_0,\rr_0,\omega} \cdot\partial_x\BB\right.\nonumber\\
&\left.+\BB\cdot\pare{\lim_{\rr_1, \rr_2\rightarrow\rr_0} {\partial_{x_1}}{\partial_{x_2}}\bbar{\ca{G}}\pare{\rr_1,\rr_2,\omega}} \cdot\BB\right].}
Here we have defined the tensor  \aleqn{&\bbar{\ca{G}}\pare{\rr_0,\rr_0,\omega}_{il}\equiv\nonumber\\
&\lim_{\rr_1,\rr_2\rightarrow\rr_0}\epsilon_{ijk}  \epsilon_{nml} \frac{\partial^2}{\partial r_{1j} \partial r_{2m}}\Im \spare{\bbar{G}_{kn}\pare{\rr_1,\rr_2,\omega}}.} Considering  the total Green's tensor as the sum of free space and scattering components $\bbar{G}\pare{\rr_1,\rr_2,\omega} = \bbar{G}^{(0)}\pare{\rr_1,\rr_2,\omega}+\bbar{G}^{(1)}\pare{\rr_1,\rr_2,\omega}$, the free space Green's tensor $\bbar{G}^{(0)}\pare{\rr_1,\rr_2,\omega}$ yields the motional decoherence rate of the diamagnet in the absence of the surface and the scattering Green's tensor $\bbar{G}^{(1)}\pare{\rr_1,\rr_2,\omega}$ gives the modifications to the decoherence due to scattering off the surface. We find that for the case of a particle trapped in a harmonic potential of frequency $\omega_{t}$ near a superconductor, the localization parameter is given by 

\aleqn{
\Lambda_B^{SC}\approx \frac{3\mu_0 \chi^2 }{64\pi\hbar }\coth\pare{\frac{\hbar\omega_t}{2K_b T}}\frac{\lambda_L^3(T)}{\delta_m^2(T)z_t^4}\left[\partial_xB_x^2\right.\nonumber\\
\left.+\partial_xB_y^2+2\partial_xB_z^2+\frac{5}{z_t^2}\pare{3B_x^2/4+B_y^2/4+B_z^2} \right]
\label{supdec}
}
with  $\BB(\rr_0) = \pare{B_x, B_y, B_z} $ as the externally applied magnetic field evaluated at the center-of-mass position of the particle, $\lambda_L(T) = m_e/\spare{\mu_0 n_s(T) e^2}$ as the London penetration length, and $\delta_m(T) = \sqrt{2n_0/\spare{\omega_t \mu_0 \sigma_m n_n(T)}}$ as the skin depth corresponding to  the normal phase of the superconductor described by the two-fluid model \cite{AnnettBook}.  Here, $m_e$ and $e$ refer to the electronic mass and charge respectively, $n_s(T)/n_t = 1- n_n(T)/n_t = 1-\pare{T/T_c}^4$ is the fractional charge carrier density of superconducting electrons, and $\sigma_m$ refers to the electrical conductivity of the metal in the normal phase. In deriving \eqnref{supdec}, we have assumed that  $\lambda_L\ll\delta_m $, and that the distance of the particle from the surface is much smaller compared to the trap wavelength $\pare{ \lambda_L\ll z_t\ll c/\omega_t}$, similar to \cite{Skagerstam2006}.   For a sphere of radius $1~\mu\text{m}$, environmental temperature $T_e=50~\text{mK}$, and $|\BB|= 1~\text{mT}$, we find that the decoherence rate for the motion of the sphere near a superconducting surface is much smaller relative to the gravitationally induced decoherence $\pare{\Lambda_{B}^{SC}/\Lambda_G\sim10^{-11}}$. Contrasting this with the decoherence near a metal surface with a skin depth $\delta_m$, such that $z_t\ll\delta_m\ll c/\omega_t$, one obtains a localization parameter that goes as
$\Lambda_{B}^{M}\sim \mu_0 \chi^2 K_b T_e \abs{\BB}^2/(\hbar^2\omega_t   z_t^3 \delta_m^2)$.  For the chosen set of parameter values as before, one has that $\Lambda_{B}^{M}/\Lambda_G \approx 10^{11} (10~\mu\text{m}/z_t)^3 (\delta_{\text{Au}}/\delta_m)^2$,
where $\delta_\text{Au}$ is the skin depth of gold.
Thus, for typical non-superconducting metals, we find that decoherence due to magnetic field fluctuations near the surface is much larger than gravitationally-induced decoherence. Therefore, to be able to observe gravitationally-induced decoherence, one needs to place the microsphere near a superconducting surface. That is, one should use a superconducting atom chip~\cite{Skagerstam2006,Nirrengarten2006,Bernon2013}.

\section{case study} \label{Sec:CaseStudy}

\begin{figure}[t]
\centering
\includegraphics[width= \columnwidth]{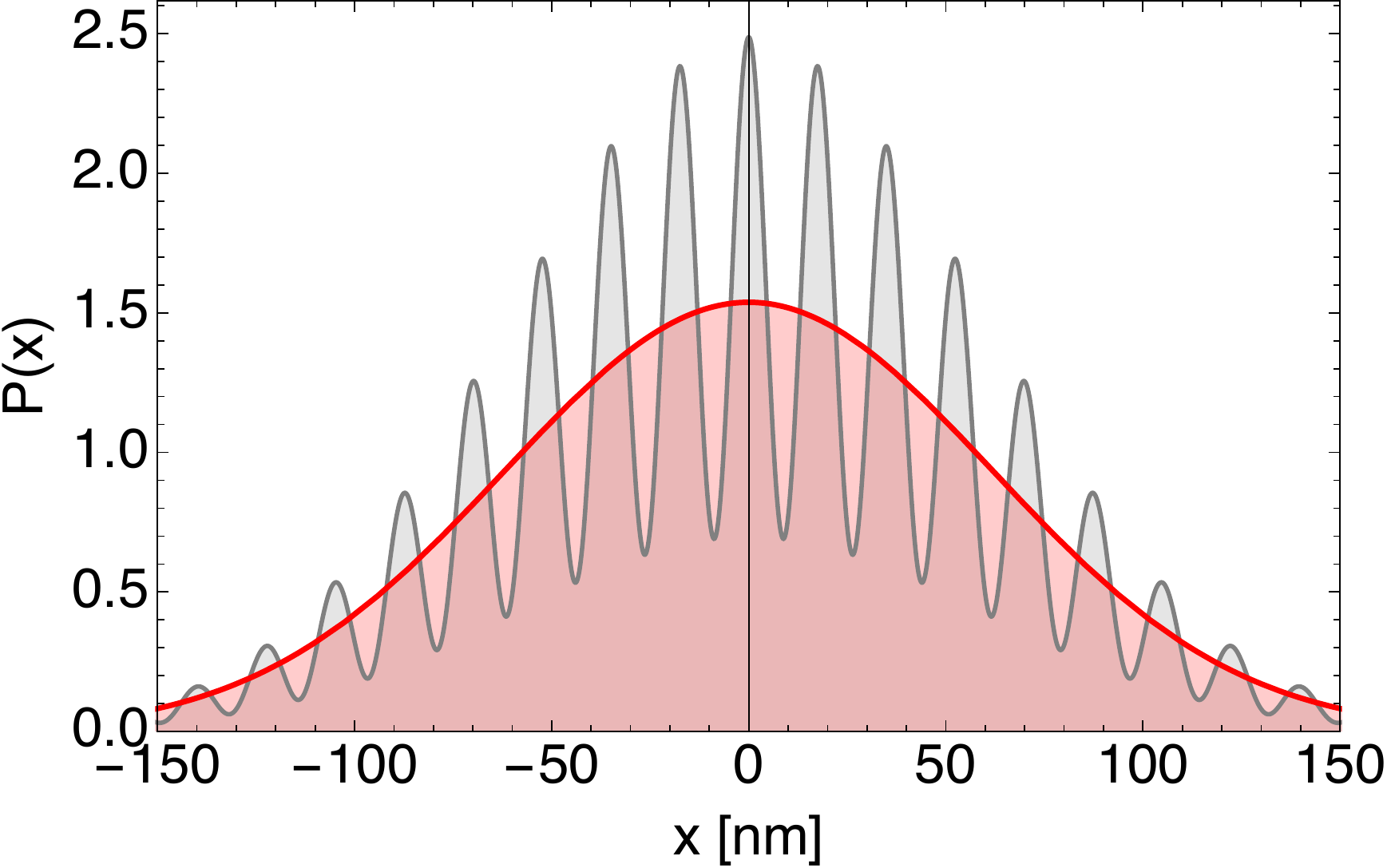}
\caption{  Interference pattern obtained for a sphere of $R=1~\mu\text{m}$ with a mass of $M\approx 2\times 10^{13}~\text{amu}$ in a Young's double slit interferometer with slit separation $d=0.5~\mu\text{m}$ and slit width $\sigma_d \approx 11.6~\text{nm}$. The whole interferometer after cooling, which takes $25~\text{ms}$, takes $596$~\text{ms}. The grey line is obtained considering standard sources of decoherence and the red one the parameter-free gravitationally-induced decoherence. In this regime, the observation of interference fringes would unambiguously falsify gravitationally-induced decoherence.}
\label{Fig5}
\end{figure}

In this section we consider a particular set of experimental parameters and analytically calculate the interference pattern that would be obtained in the proposed quantum micromechanical interferometer. This is done considering standard sources of decoherence and with (red curve in \figref{Fig5}) or without (grey curve in \figref{Fig5}) gravitationally-induced decoherence. With the parameters explicitly listed in the Appendix, one obtains the interference pattern plot in \figref{Fig5}, indicating that the faint parameter-free gravitationally-induced decoherence could be unambiguously falsified.  We emphasize there would be other sets of parameters that would lead to the same conclusion, namely that gravitationally-induced decoherence could be falsified.  

We have considered a superconducting sphere in the Meissner state with a radius of $R=1~\mu\text{m}$ and a mass of $M\approx2\times10^{13}~\text{amu}$, well inside the gravitational quantum regime, as shown in \figref{Fig1}. The superposition size is determined by the slit separation $d=0.5~\mu\text{m}$ and slit width $\sigma_d \approx 11.6~\text{nm}$. The total time of the quantum micromechanical interferometer after cooling (which takes $t_1=25~\text{ms}$) is $T=596~\text{ms}$. This time is distributed between the different steps as $t_2=17~\text{ms}$, $t_3=483~\text{ms}$, $t_4=12.7~\text{ms}$, $t_5=2.5~\text{ms}$, $t_6=61~\text{ms}$, and $t_7=20~\text{ms}$. The harmonic potentials in steps 2, 5, and 7 have a frequency of $50~\text{Hz}$. Note that the longest duration in the operation of the interferometer is in the free expansion of step 3, where standard sources of decoherence are minimal. The coherence length after step 3 is given by $\xi(T_3)\approx 607~\text{nm}$ without gravitationally-induced decoherence, and $\xi_G(T_3)\approx 121~\text{nm}$ including it. Hence note that the condition $\xi_G \ll d \lesssim \xi$ is fulfilled. The latter is the key condition in order to falsify gravitationally-induced decoherence, as shown in \figref{Fig5}. Steps 4, 5, 6, and 7 are intended to build the interference pattern, in case of a coherent superposition, and thus have to be performed as fast as possible in order to prevent decoherence from washing-out the interference pattern. With the parameters used, the blurring distance is given by $\sigma_\Lambda\approx2.4~\text{nm}$ with the presence of gravitationally-induced decoherence and $\sigma_\Lambda\approx1.9~\text{nm}$ with only standard sources of decoherence. Note that the difference is minimal and the blurring is small enough to not wash-out the interference fringes generated when gravitationally-induced decoherence is absent, since the fringe separation takes the value $x_f \approx 24.7~\text{nm}$ for the chosen parameters.

A larger mass of the sphere has not been considered mainly because of two reasons: (i) the magnetic field gradients are limited by the fact that the field at the surface of the sphere has to be smaller than the first critical field of the superconducting material. The larger the mass, the smaller the field gradients and thus the smaller the trap frequency in step 1, which then prevents one from achieving the resolved-sideband regime and thus preparing a sufficiently pure initial state. Namely, we require the initial thermal mean phonon number occupation to be smaller than 1, and  $\bar n \approx 0.07$ for our parameter choice. (ii) The larger the mass, the more vibration isolation is required to ensure that gravitationally-induced decoherence is the main source of decoherence. We have already considered a very good isolation vibration with values of $\sqrt{S_{xx}} \sim 10^{-16}~\text{m}/\sqrt{\text{Hz}}$ at frequencies below $1~\text{Hz}$, as required in steps 3, 4, and 7, and $\sqrt{S_{xx}}\sim 10^{-18}~\text{m}/\sqrt{\text{Hz}}$ at frequencies of $50~\text{Hz}$, as required in steps 2,5 and 6. In a cryogenic environment these values are challenging, and comparable to vibration isolation in present-day gravitational wave detectors. Nevertheless, we remark that in future cryogenic gravitational wave detectors, schemes for achieving vibration isolation at the incredible numbers of $10^{-24}~\text{m}/\sqrt{\text{Hz}}$ at low frequencies are being developed, see for instance the KAGRA project~\cite{kagra}.

Vibrations are the main source of decoherence in our model provided that the decoherence from magnetic field fluctuations can be ignored. The latter is the case if the surface is superconducting, for example, if one uses a superconducting atom chip~\cite{Skagerstam2006,Nirrengarten2006,Bernon2013}. Note that the sphere is skating at a distance from the surface of the order of $11~\mu\text{m}$. The distance slightly varies in each step of the protocol, something that we have not optimized since the smooth transitions between steps have not been included in our calculations. Should the decoherence from the surface be a hindering factor, one could also consider implementing the interferometer protocol in free fall~\cite{Cronin2009} with external magnetic potentials speeding-up the dynamics, as in step 2 and 6 of the protocol. 

As mentioned before, we refer to the Appendix for the complete list of experimental parameters used to analytically calculate, as an example, the interference pattern shown in \figref{Fig5}.

\section{Conclusions \& Final Remarks} \label{Sec:Conclusions}

We have proposed and analyzed a quantum micromechanical interferometer scheme for preparing and probing quantum superpositions of a sphere in the micrometer scale with a mass $\gtrsim 10^{13}~\text{amu}$.  The quantum superposition state is prepared by using a Young's double slit with slit separation comparable to the radius of the sphere. In particular, we have proposed an all-magnetic on-chip scheme where a superconducting microsphere skates in a magnetic landscape generated by persistent currents. The chip also houses some SQUIDs that are used to cool, implement a double slit, and measure final interference fringes. As the sphere skates through such a magnetic skatepark an all-magnetic Young's double slit experiment is implemented passively, namely without the need to switch on and off magnetic potentials and to actively control the quantum circuits. 

Using a challenging but feasible set of experimental parameters  as a case study, we have shown that one can enter into the gravitational quantum regime where the faint parameter free gravitationally-induced collapse model, proposed by Di\'osi and Penrose~\cite{Diosi1984,Penrose1996}, could be unambiguously falsified. While we have focused on this paradigmatic parameter-free collapse model as a figure of merit, we remark that our proposal would falsify the continuous-spontaneous-localization collapse model~\cite{Bassi2013} with a collapse frequency phenomenological parameter as small as $10^{-22}~\text{Hz}$, several orders of magnitude below the original value proposed by Ghirardi, Rimini, and Weber in the 80's~\cite{Ghirardi1986}.

In our analysis, especially regarding the experimental feasibility, we have modelled the experiment with some necessary degree of simplification. For example, we have not considered the many potential sources of noise and decoherence that could be encountered in a real experimental implementation. Instead, we have analyzed the aspects that we consider inevitable in any experimental implementation and that would  pose more than a technological challenge. The on-chip implementation of the quantum micromechanical interferometer is ambitious and therefore experimentally challenging, specially when operating in the regime presented in the case study. We suggest to address this goal from a top-down approach, namely to gradually increase the delocalization distance of a superconducting microsphere brought to the quantum regime. To do so, one could perform a set of progressively challenging experiments that could be themselves relevant in the the field of quantum nano- and micromechanics~\cite{revopt}. A possible experimental roadmap could be to show: (i) Magnetic levitation of a superconducting microsphere in vacuum,  (ii) Magnetomechanical coupling between a superconducting object and a quantum circuit (the superconducting object does not need to be levitating, see for instance~\cite{Via2015}), (iii) Ground state cooling of a magnetically levitated superconducting sphere using a quantum magnetomechanical coupling by combining the two previous points~\cite{RomeroIsart2012}, (iv) The preparation of non-trivial quantum states of the center of mass of a magnetically levitated superconducting microsphere using a quantum magnetomechanical coupling to a quantum circuit~\cite{RomeroIsart2012} (these states still have a small delocalization distance comparable to the zero point motion), (v)  Quantum coherent expansion of the center-of-mass motion by making the trap shallower after ground state cooling, (vi) Exponential speed-up of the coherent expansion by inverting the quadratic potential after ground state cooling, (vii) The implementation of ground state cooling and coherent expansion as proposed in step 1, 2, and 3 of quantum micromechanical interferometer, (viii) Perform a squared continuous-time quantum measurement using the coupling to a quantum circuit to prepare a double slit, and (xix) The implementation of the quantum micromechanical interferometer with the 7 steps with small slit separations (one can then gradually increase the slit separation distance).
Note that this roadmap would also allow to carefuly characterize the standard sources of decoherence arising from scattering of gas molecules, vibrations, and magnetic field fluctuations, as well as other sources of decoherence that may have been overlooked.

From a broader perspective, we think that experiments aiming to explore the quantum superposition principle of very massive objects would benefit from the conditions that we have analysed and determined to be crucial for this proposal: (i) a cryogenic environment that minimizes decoherence due to black-body radiation and due to scattering of gas molecules because of the ultra-high vacuum present in such environments. We note, for example, that pressures less than $\times~10^{-17}$ mbar have been achieved in cryogenic environments \cite{Gabrielse90,GabrielseRev}, (ii) the possibility to use static fields (ideally using persistent currents) to generate inverted potentials that speed-up the otherwise very slow quantum dynamics, (iii) a very good vibration isolation in a cryogenic environment to avoid decoherence when using external potentials, (iv) superconducting surfaces in case of near-surface proposals since low frequency electromagnetic field fluctuations are absent, and (v) the use of quantum circuits to cool, to engineer non-linearities, and to measure the position of the massive object. 

We remark that one of the main ingredients to perform quantum interference of such massive objects in less than a second is the use of inverted potentials that speed-up the quantum dynamics, as justified in more detail in~\cite{RomeroIsart2016}. This idea could also be implemented in a free fall experiment that would require a cryogenic high-vacuum tower of few meters~\cite{Cronin2009}, employing the same protocol as described here. In our protocol we propose to perform the Young's double slit by performing a quantum measurement of the squared position. Alternatively, one could also consider preparing the quantum superposition using a non-Gaussian (more than quadratic dependence in position) magnetic potential. Also, one could consider a magnetic grating formed, for instance, using the field created by an array of pinned superconducting vortices or nanomagnets. If the magnetic lattice has field maxima larger than the critical field of the sphere, superconductivity is prevented for spheres passing through these areas. Such a magnetic grating could be used for near-field matter-wave interferometry, as an all-magnetic version of the all-optical Talbot-Lau matter-wave interferometers currently being implemented~\cite{Haslinger2013}. In this all-magnetic scenario, one could again use static fields to speed-up the dynamics without creating decoherence and thus increase the mass of the particle used in near-field matter-wave interferometers without requiring a space environment~\cite{MAQRO,Kaltenbaek2015}.
Our proposal could also be extended to exploit the other interpretations of the gravitational quantum regime. In particular, one could consider two cold superconducting spheres moving along parallel magnetic waveguides and measure the mutual gravitational interaction which can then be used to determine $G$. That is a levitation version of a recently proposed experiment for measuring $G$ with milligram masses~\cite{Schmoele2016}. 

We hope that the proposal presented in this manuscript, namely an earth-based on-chip all-magnetic quantum micromechanical interferometer exploring quantum mechanics in a hitherto unexplored parameter regime, will trigger further table-top experimental and theoretical research on the interplay between quantum mechanics and gravity.

 This work is supported by the European Research Council (ERC-2013-StG 335489 QSuperMag) and the Austrian Federal Ministry of Science, Research, and Economy (BMWFW). ORI acknowledges inspiring discussions with M. Aspelmeyer, J. I. Cirac, and S. Dimopoulos.

\appendix

\section{Experimental parameters} \label{Sec:ExpParam}
In this section we provide a list of experimental parameters. We emphasize that the choice of parameters is not optimized, it is simply a representative example. This set of parameters has been used to plot \figref{Fig5}. We label the free parameters that we have chosen with the item symbol $\star$. The parameters that are not free, either because they are calculated from the chosen free parameters or because they are fixed by the choice (\eg~mass density once we have chosen the material), are labeled with the item symbol $\bullet$. 

\subsection{Superconducting sphere}

\begin{itemize}
\item[$\star$] Material: Niobium.
\item Density $\rho = 8570~\text{Kg}/\text{m}^3$.
\item First critical field $140~\text{mT}$.
\item Critical temperature at zero field $9.2~\text{K}$. 
\item[$\star$] Radius $R=1~\mu\text{m}$.
\item Mass $M\approx 2 \times 10^{13}~\text{amu}$.
\end{itemize}

\subsection{Environmental conditions}
\begin{itemize}
\item[$\star$] Bulk and environmental temperature $T_e= T_i = 50~\text{mK}$.
\item[$\star$] Pressure $P = 10^{-17}~\text{mbar}$~\cite{Gabrielse90,GabrielseRev}.
\item[$\star$] Vibrations: $\sqrt{S_{xx}} = 10^{-16}~\text{m}/\sqrt{\text{Hz}}$ at $\lesssim 1~\text{Hz}$ and $\sqrt{S_{xx}} = 10^{-18}~\text{m}/\sqrt{\text{Hz}}$ at $50~\text{Hz}$.
\end{itemize}

\subsection{Decoherence}
\begin{itemize}
\item Scattering of gas molecules $1/\gamma_a= 1.6~\text{s}$ (assuming He).
\item Black-body decoherence due to scattering is
\be
\frac{\Lambda^s_{bb}}{\Lambda_G} \approx 9.4 \times 10^{-27} \times \spare{\frac{\epsilon(\w_\text{th}) - 1}{\epsilon(\w_\text{th}) +2 }}^2
\ee
\item Black-body decoherence due to absorption and emission is
\be
\frac{\Lambda_{bb}^{e(a)}}{\Lambda_G} \approx 2.1 \times 10^{-15} \times \text{Im} \spare{\frac{\epsilon(\w_\text{th}) - 1}{\epsilon(\w_\text{th}) +2 }}
\ee
\item Localization parameter due to vibrations in a Harmonic potential of frequency $\w$ is
\be 
\frac{\Lambda_f}{\Lambda_G} \approx 1.4 \times 10^{10} \times \frac{\w^4 S_{xx}(\w)}{\text{m}^2 \text{Hz}^3} 
\ee 

\end{itemize}

\subsection{Magnetic skatepark}

\subsubsection{Wires}

\begin{itemize}
\item[$\star$] Current interior wires $I_{\rm i}=1.00~\text{A}$.
\item[$\star$] Current outer wires $I_{\rm o}=0.92~\text{A}$.
\item[$\star$] Distance interior wires $d_{\rm i}=13R$.
\end{itemize}

\subsubsection{Pick-up coils}

\begin{itemize}
\item[$\star$] Position pick-up coil in step 1: $x_c=3 R$  and $z_c=0$.
\item[$\star$] Size pick-up coil in step 1: $l_x=6 R$.
\item[$\star$] Position pick-up coil in step 4 and 7:  $x_c=z_c=0$.
\item[$\star$] Size pick-up coil in step 4 and 7: $l_x=5 R$.
\end{itemize}

\subsection{Protocol}

\subsubsection{Step 1: Cooling}

\begin{itemize}
\item [$\star$] Distance outer wires $d_o = 30 R$
\item Trap distance $z_t = 10.923 R$ and frequency $\w_z/(2 \pi)= 2193.9 ~\text{Hz}$ along the $z$-axis.
\item Attractive potential frequency $\w_1/(2 \pi)= 2159.1 ~\text{Hz}$.
\item[$\star$] Cavity frequency $\w_{c_1}/(2 \pi)=1~\text{GHz}$.
\item[$\star$] Cavity decay rate $\kappa_{1}/(2 \pi)=1000~\text{Hz}$.
\item Constant flux coefficient $s_0=0.208$.
\item Magnetomechanical coefficient $\eta_l/\sigma =-1.13 \times 10^{-5} \; \text{nm}^{-1}$.
\item Linear magnetomechanical single-photon coupling $g_{l}/(2 \pi)=1.8~\text{Hz}$.
\item[$\star$] Intra-cavity mean-field photon number $|\alpha_1|=200$.
\item[$\star$] Detuning of the driving field $\Delta=-\w_1$.
\item Adiabatic elimination condition $g_l \abs{\alpha_1}/(2\kappa_1) = 0.17$.
\item[$\star$] Cooling time $t_1=25~\text{ms}$.
\item[$\star$]  Initial mean number of phonons after feedback cooling $n_0=1000$.
\item  Final mean number of phonons $\bar n \approx 0.057$.
\item Coherence length $\xi(0)\approx 8.8 \times 10^{-13}~\text{m}$.
\end{itemize}

\subsubsection{Step 2: Boost}

\begin{itemize}
\item [$\star$] Distance outer wires $d_o = 14.3408 R$
\item Trap distance $z_t = 11.852 R$ and frequency $\w_z/(2 \pi)= 394.2~\text{Hz}$ along the $z$-axis.
\item Inverted potential frequency $\w_2/(2 \pi)=50.0~\text{Hz}$.
\item[$\star$] Boosting time $t_2=17~\text{ms}$.
\item Momentum gain $\sqrt{v_p(t_2)/v_p(0)}\approx104$.
\item Standard vs gravitational-induced decoherence $(\Lambda_2-\Lambda_G)/\Lambda_G\approx1.4$.
\item Position variance without $\Lambda_G$: $\sqrt{v_x(t_2)}\approx 1.6 ~\text{nm}$.
\item Coherence length without $\Lambda_G$: $\xi(t_2)\approx 4.0~\text{nm}$.
\item Coherence length with $\Lambda_G$: $\xi(t_2)\approx 4.0~\text{nm}$.
\end{itemize}

\subsubsection{Step 3: Free}

\begin{itemize}
\item [$\star$] Distance outer wires $d_o \approx 14.388 R$
\item Trap distance $z_t = 11.787 R$ and frequency $\w_z/(2 \pi)= 400.5~\text{Hz}$ along the $z$-axis.
\item Attractive potential frequency $\w_3/(2 \pi) \approx 0~\text{Hz}$.
\item[$\star$] Free time $t_3=483~\text{ms}$.
\item Position variance without $\Lambda_G$: $\sqrt{v_x(T_3)}\approx239~\text{nm}$.
\item Position variance with $\Lambda_G$: $\sqrt{v_x(T_3)}\approx239~\text{nm}$.
\item Coherence length without $\Lambda_G$: $\xi(T_3)\approx607~\text{nm}$.
\item Coherence length with $\Lambda_G$: $\xi(T_3)\approx121~\text{nm}$.
\end{itemize}

\subsubsection{Step 4: Split}

\begin{itemize}

\item [$\star$] Distance outer wires $d_o \approx 14.388 R$
\item Trap distance $z_t = 11.787 R$ and frequency $\w_z/(2 \pi)= 400.5~\text{Hz}$ along the $z$-axis.
\item Length unit $\sigma_4=\xi(T_3)\approx607~\text{nm}$.
\item[$\star$] Cavity frequency $\w_{c_4}/(2 \pi)=1~\text{GHz}$.
\item[$\star$] Cavity decay rate $\kappa_{4}/(2 \pi)=1~\text{GHz}$.
\item Constant flux coefficient $s_0=0.520$.
\item  Magnetomechanical coefficient $\eta_q/\sigma^2 =-1.09 \times 10^{-10} \; \text{nm}^{-2}$.
\item Quadratic magnetomechanical single-photon coupling $g_{q}/(2 \pi)\approx25098~\text{Hz}$.
\item[$\star$] Intra-cavity mean-field photon number $|\alpha_4|=100$.
\item Adiabatic elimination condition $g_q \abs{\alpha_4}/(2\kappa_4) \approx 0.001$.
\item[$\star$] Measurement time $t_4=12.7~\text{ms}$.
\item Initial phase $\Theta \approx 2.00 \times 10^7$.
\item Final phase $F(t_4)\approx 0$.
\item Measurement strength $\chi_4 = 31.7$.
\item [$\star$]  Slit separation $d=500~\text{nm}$.
\item Slit width $\sigma_d=11.63~\text{nm}$.

\end{itemize}

\subsubsection{Step 5: Rotation}

\begin{itemize}

\item [$\star$] Distance outer wires $d_o = 14.431 R$
\item Trap distance $z_t = 11.729 R$ and frequency $\w_z/(2 \pi)=406.3~\text{Hz}$ along the $z$-axis.
\item Attractive potential frequency $\w_5/(2 \pi)=50.0~\text{Hz}$.
\item Rotation time $t_5= \pi/(4 \w_5) =2.5~\text{ms}$.
\item Standard vs gravitational-induced decoherence $(\Lambda_5-\Lambda_G)/\Lambda_G\approx1.4$.
\item Convolution factors that justify neglecting decoherence $A_1\approx 10^{-19}$ and $A_2\approx 10^{-4}$.

\end{itemize}

\subsubsection{Step 6: Inflation}

\begin{itemize}

\item [$\star$]  Distance outer wires $d_o = 14.3408 R$
\item  Trap distance $z_t = 11.852 R$ and frequency $\w_z/(2 \pi)= 394.2~\text{Hz}$ along the $z$-axis.
\item Inverted potential frequency $\w_6/(2 \pi)=50.0~\text{Hz}$.
\item[$\star$] Inflation time $t_6=61~\text{ms}$.
\item Standard vs gravitational-induced decoherence $(\Lambda_6-\Lambda_G)/\Lambda_G\approx1.4$.
\item Blurring coefficient without $\Lambda_G$:  $\sigma_\Lambda \approx 1.9~\text{nm}$.
\item Blurring coefficient with $\Lambda_G$:  $\sigma_\Lambda \approx 2.4~\text{nm}$.
\item Fringe separation at $T_6$: $24.7~\text{nm}$

\end{itemize}

\subsubsection{Step 7: Measure}

\begin{itemize}
\item [$\star$] Distance outer wires $d_o \approx 14.388 R$
\item Trap distance $z_t = 11.787 R$ and frequency $\w_z/(2 \pi)= 400.5~\text{Hz}$ along the $z$-axis.
\item Length unit without $\Lambda_G$: $\sigma_7\approx 59.9~\text{nm}$.
\item Length unit with $\Lambda_G$: $\sigma_7\approx62.0~\text{nm}$.
\item[$\star$] Cavity frequency $\w_{c_7}/(2 \pi)=1~\text{GHz}$.
\item[$\star$] Cavity decay rate $\kappa_{7}/(2 \pi)=1\times 10^{6}~\text{Hz}$.
\item Constant flux coefficient $s_0=0.520$.
\item  Magnetomechanical coefficient $\eta_q/\sigma^2 =-1.09 \times 10^{-10} \; \text{nm}^{-2}$.
\item Quadratic magnetomechanical single-photon coupling $g_{q}/(2 \pi)\approx262~\text{Hz}$.
\item[$\star$] Intra-cavity mean-field photon number $|\alpha_7|=500$.
\item Adiabatic elimination condition $g_q \abs{\alpha_7}/(2\kappa_7) = 0.07$.
\item[$\star$] Measurement time $t_7=20~\text{ms}$.
\item Measurement strength $\chi_7 = 65.7$.
\item Position resolution $\delta x \approx 5~\text{nm}$.
\item Total protocol time after cooling $T_7=596~\text{ms}$.
\end{itemize}


\begin{thebibliography}{27}



\expandafter\ifx\csname natexlab\endcsname\relax\def\natexlab#1{#1}\fi
\expandafter\ifx\csname bibnamefont\endcsname\relax
  \def\bibnamefont#1{#1}\fi
\expandafter\ifx\csname bibfnamefont\endcsname\relax
  \def\bibfnamefont#1{#1}\fi
\expandafter\ifx\csname citenamefont\endcsname\relax
  \def\citenamefont#1{#1}\fi
\expandafter\ifx\csname url\endcsname\relax
  \def\url#1{\texttt{#1}}\fi
\expandafter\ifx\csname urlprefix\endcsname\relax\def\urlprefix{URL }\fi
\providecommand{\bibinfo}[2]{#2}
\providecommand{\eprint}[2][]{\url{#2}}


\bibitem{Arndt1999}
M.~Arndt, O.~Nairz, J.~Vos-Andreae, C.~Keller, G.~van~der~Zouw, and A.~Zeilinger, {\em Wave–particle duality of C60 molecules}, \href{http://www.nature.com/nature/journal/v401/n6754/abs/401680a0.html}{Nature~(London) {\bf 401}, 680 (1999).}

\bibitem{Hornberger2012}
K.~Hornberger, S.~Gerlich, P.~Haslinger, S.~Nimmrichter, and M.~Arndt, {\em 
Colloquium: Quantum interference of clusters and molecules}, \href{http://journals.aps.org/rmp/abstract/10.1103/RevModPhys.84.157}{Rev.~Mod.~Phys. {\bf 84}, 157 (2012).}

\bibitem{Cronin2009}
A.~D.~Cronin, J.~Schmiedmayer, and D.~E.~Pritchard, {\em Optics and interferometry with atoms and molecules}, \href{http://journals.aps.org/rmp/abstract/10.1103/RevModPhys.81.1051}{Rev.~Mod.~Phys. {\bf 81}, 1051 (2009).}

\bibitem{Arndt2014}
M.~Arndt and K.~Hornberger, {\em Testing the limits of quantum mechanical superpositions}, \href{http://www.nature.com/nphys/journal/v10/n4/full/nphys2863.html}{Nat.~Phys. {\bf 10}, 271 (2014).}

\bibitem{Eibenberger2013}
S.~Eibenberger, S.~Gerlich, M.~Arndt, M.~Mayor, and J.~T\"uxen, {\em Matter–wave interference of particles selected from a molecular library with masses exceeding 10 000 amu}, \href{http://pubs.rsc.org/en/Content/ArticleLanding/2013/CP/c3cp51500a#!divAbstract}{Phys.~Chem.~Chem.~Phys. {\bf 15}, 14696 (2013).}

\bibitem{Footnote1}
The gravitationally-induced collapse model introduced by Di\'osi and Penrose is parameter independent when an homogeneous mass density is assumed. To avoid divergences at the microscopic level, a cutoff distance $r_0$ in the mass density is typically introduced, thereby making the model parameter-dependent. In that case the quantum superposition lifetime for a sphere of radius $R$ separated by a distance of the order $R$ would then be reduced by a factor $r_0/R$, namely it would be given by $\tau_G r_0/R$. Thus the lifetime $\tau_G$ discussed in the text is a parameter-free upper-bound to the lifetime predicted by gravitationally-induced collapse models. See \cite{Diosi2007,ORI2011b} for further details.


\bibitem{Diosi2007}
L.~Di\'osi, {\em Notes on certain Newton gravity mechanisms of wavefunction localization and decoherence}, \href{http://iopscience.iop.org/article/10.1088/1751-8113/40/12/S07/meta;jsessionid=6A3176BC7BF382793F5870BCAAEF0DA7.c4.iopscience.cld.iop.org}{J.~Phys.~A:~Math.~Theor. {\bf 40}, 2989 (2007)}.

\bibitem{ORI2011b}
O.~Romero-Isart, {\em Quantum superposition of massive objects and collapse models}, \href{http://journals.aps.org/pra/abstract/10.1103/PhysRevA.84.052121}{Phys.~Rev.~A {\bf 84}, 052121 (2011).}


\bibitem{Diosi1984}
L.~Di\'{o}si, {\em Gravitation and quantum-mechanical localization of macro-objects}, \href{http://www.sciencedirect.com/science/article/pii/0375960184903979}{Phys.~Lett.~A {\bf 105}, 199 (1984).}

\bibitem{Penrose1996}
R.~Penrose, {\em On Gravity's role in Quantum State Reduction}, \href{http://link.springer.com/article/10.1007/BF02105068}{Gen.~Relativ.~Gravit.~\textbf{28}, 581 (1996).}

\bibitem{Schmoele2016}
J.~Schmöle, M.~Dragosits, H.~Hepach, M.~Aspelmeyer, {\em A micromechanical proof-of-principle experiment for measuring the gravitational force of milligram masses}, \href{http://arxiv.org/abs/1602.07539}{arXiv:1602.07539.}

\bibitem{Zeh2011}
H.~D.~Zeh, {\em Feynman’s interpretation of quantum theory}, \href{http://link.springer.com/article/10.1140/epjh/e2011-10035-2}{Eur.~Phys.~J.~H {\bf 36}, 63 (2011).}

\bibitem{Anastopoulos2015}
C.~Anastopoulos and B.~L.~Hu, {\em Probing a gravitational cat state}, \href{http://iopscience.iop.org/article/10.1088/0264-9381/32/16/165022/meta;jsessionid=A483988D12959A94FBD046EF68E1F306.c2.iopscience.cld.iop.org}{Classical~Quant.~Grav. {\bf 32}, 165022 (2015).}

\bibitem{oconell10}
A.~D.~O’Connell \etal, {\em Quantum ground state and single-phonon control of a mechanical resonator}, \href{http://www.nature.com/nature/journal/v464/n7289/abs/nature08967.html}{Nature~(London) {\bf 464}, 697 (2010).}

\bibitem{teufel11}
J.~D.~Teufel,	T.~Donner,	D.~Li,	J.~W.~Harlow,	M.~S.~Allman,	K.~Cicak,	A.~J.~Sirois,	J.~D.~Whittaker,	K.~W.~Lehnert, and R.~W.~Simmonds, {\em Sideband cooling of micromechanical motion to the quantum ground state}, \href{http://www.nature.com/nature/journal/v475/n7356/full/nature10261.html}{Nature~(London) {\bf 475}, 359 (2011).}

\bibitem{chan11}
J.~Chan,	T.~P.~Mayer~Alegre,	A.~H.~Safavi-Naeini,	J.~T.~Hill,	A.~Krause,	S.~Gr\"oblacher,	M.~Aspelmeyer, and O.~Painter, {\em Laser cooling of a nanomechanical oscillator into its quantum ground state}, \href{http://www.nature.com/nature/journal/v478/n7367/abs/nature10461.html}{Nature~(London) {\bf 478}, 89 (2011).}

\bibitem{revopt}
M.~Aspelmeyer, T.~J.~Kippenberg, and F.~Marquardt, {\em Cavity optomechanics}, \href{http://journals.aps.org/rmp/abstract/10.1103/RevModPhys.86.1391}{Rev.~Mod.~Phys {\bf 86}, 1391 (2014).}

\bibitem{ORI2010}
O.~Romero-Isart, M.~L.~Juan, R.~Quidant, and J.~I.~Cirac, {\em Toward quantum superposition of living organisms}, \href{http://iopscience.iop.org/article/10.1088/1367-2630/12/3/033015/meta}{New~J.~Phys. {\bf 12}, 033015 (2010).}

\bibitem{Chang2010}
D.~E.~Chang, C.~A.~Regal, S.~B.~Papp, D.~J.~Wilson, J.~Ye, O.~Painter, H.~J.~Kimble, and P.~Zoller, {\em Cavity opto-mechanics using an optically levitated nanosphere}, \href{http://www.pnas.org/content/107/3/1005}{Proc.~Nat.~Acad.~Sci.~U.~S.~A. {\bf 107}, 1005 (2010).}

\bibitem{Barker2010}
P.~F.~Barker and M.~N.~Shneider, {\em Cavity cooling of an optically trapped nanoparticle}, \href{http://journals.aps.org/pra/abstract/10.1103/PhysRevA.81.023826}{Phys.~Rev.~A {\bf 81}, 023826 (2010).}

\bibitem{Kiesel2013}
N.~Kiesel, F.~Blaser, U.~Delić, D.~Grass, R.~Kaltenbaek, and M.~Aspelmeyer, {\em Cavity cooling of an optically levitated submicron particle}, \href{http://www.pnas.org/content/110/35/14180.abstract}{Proc.~Nat.~Acad.~Sci.~U.~S.~A. {\bf 110}, 14180 (2013).}

\bibitem{Asenbaum2013}
P.~Asenbaum,	S.~Kuhn, S.~Nimmrichter,	U.~Sezer,	and M.~Arndt, {\em Cavity cooling of free silicon nanoparticles in high vacuum}, \href{http://www.nature.com/ncomms/2013/131106/ncomms3743/full/ncomms3743.html}{Nat.~Commun. {\bf 4}, 2743 (2013).}



\bibitem{Millen2015}
J.~Millen, P.~Z.~G.~Fonseca, T.~Mavrogordatos, T.~S.~Monteiro, and P.~F.~Barker, {\em Cavity Cooling a Single Charged Levitated Nanosphere}, \href{http://journals.aps.org/prl/abstract/10.1103/PhysRevLett.114.123602}{Phys.~Rev.~Lett. {\bf 114}, 123602 (2015).}

\bibitem{Gieseler2012}
J.~Gieseler, B.~Deutsch, R.~Quidant, and L.~Novotny, {\em Subkelvin Parametric Feedback Cooling of a Laser-Trapped Nanoparticle}, \href{http://journals.aps.org/prl/abstract/10.1103/PhysRevLett.109.103603}{Phys.~Rev.~Lett. {\bf 109}, 103603 (2012).}

\bibitem{ORI2011}
O.~Romero-Isart, A.~C.~Pflanzer, F.~Blaser, R.~Kaltenbaek, N.~Kiesel, M.~Aspelmeyer, and J.~I.~Cirac, {\em Large Quantum Superpositions and Interference of Massive Nanometer-Sized Objects}, \href{http://journals.aps.org/prl/abstract/10.1103/PhysRevLett.107.020405}{Phys.~Rev.~Lett. {\bf 107}, 020405 (2011).}



\bibitem{Bateman2014}
J.~Bateman,	S.~Nimmrichter,	K.~Hornberger, and H.~Ulbricht, {\em Near-field interferometry of a free-falling nanoparticle from a point-like source}, \href{http://www.nature.com/ncomms/2014/140902/ncomms5788/full/ncomms5788.html}{Nat.~Commun. {\bf 5}, 4788 (2014).}

\bibitem{MAQRO}
R.~Kaltenbaek, G.~Hechenblaikner, N.~Kiesel, O.~Romero-Isart, K.~C.~Schwab, U.~Johann, and M.~Aspelmeyer, {\em Macroscopic quantum resonators (MAQRO)}, \href{http://link.springer.com/article/10.1007\%2Fs10686-012-9292-3}{Exp.~Astron. \textbf{34}, 123~(2012).}

\bibitem{Kaltenbaek2015}
R.~Kaltenbaek \etal, {\em Macroscopic quantum resonators (MAQRO): 2015 Update}, \href{http://arxiv.org/abs/1503.02640}{arXiv:1503.02640.}

\bibitem{RomeroIsart2012}
O.~Romero-Isart, L.~Clemente, C.~Navau, A.~Sanchez, and J.~I.~Cirac, {\em Quantum Magnetomechanics with Levitating Superconducting Microspheres}, \href{http://journals.aps.org/prl/abstract/10.1103/PhysRevLett.109.147205}{Phys.~Rev.~Lett. {\bf 109}, 147205 (2012).}

\bibitem{RomeroIsart2016}
O.~Romero-Isart, {\em Coherent Inflation for Large Quantum Superpositions of Microspheres}, \href{https://arxiv.org/abs/1612.04290}{arXiv:1612.04290}.






\bibitem{Gabrielse90} G.~Gabrielse, X.~Fei,  L.~A.~Orozco, R.~L.~Tjoelker,  J.~Haas, H.~Kalinowsky, T.~A.~Trainor, and W.~Kells, {\em Thousandfold  Improvement in the Measured Antiproton Mass}, \href{http://journals.aps.org/prl/abstract/10.1103/PhysRevLett.65.1317}{Phys.~Rev.~Lett. {\bf 65}, 1317 (1990).}

\bibitem{GabrielseRev} G.~Gabrielse, { \em Comparing the antiproton and proton, and opening the way to cold hydrogen}, \href{http://www.sciencedirect.com/science/article/pii/S1049250X01800379}{Advances in Atomic, Molecular, and Optical Physics \textbf{45}, 1 (2001).}

\bibitem{Joos2003}
E.~Joos, H.~D.~Zeh, C.~Kiefer, D.~Giulini, J.~Kupsch, and I.~-O.~Smatescu, {\em Decoherence and the Appearance of a Classical World in Quantum Theory} (Springer, New York, 2003).

\bibitem{Schlosshauer2007}
M.~A.~Schlosshauer, {\em Decoherence and the Quantum-to-Classical Transition} (Springer, Berlin, 2007).



\bibitem{Henkel1999}
C.~Henkel, S.~P\"otting, M.~Wilkens {\em Loss and heating of particles in small and noisy traps}, \href{http://link.springer.com/article/10.1007/s003400050823}{App.~Phys.~B {\bf 69}, 379 (1999).}

\bibitem{Breuer2002}
H.~-P.~Breuer and F.~Petruccione, {\em The Theory of Open Quantum Systems} (Oxford University Press, Oxford, 2002).

\bibitem{Steck2014}
D.~A.~Steck, {\em Quantum and Atom Optics.} Available online at \url{http://steck.us/teaching} (2014).

\bibitem{WilsonRae07}
I.~ Wilson-Rae, N.~ Nooshi, W.~ Zwerger, and T.~J.~Kippenberg, {\em Theory of Ground State Cooling of a Mechanical Oscillator Using Dynamical Backaction}, \href{http://journals.aps.org/prl/abstract/10.1103/PhysRevLett.99.093901}{Phys.~Rev.~Lett. {\bf 99}, 093901 (2007).}

\bibitem{Marquardt07}
F.~Marquardt, J.~P.~Chen, A.~A.~Clerk, and S.~M.~Girvin, {\em Quantum Theory of Cavity-Assisted Sideband Cooling of Mechanical Motion}, \href{http://journals.aps.org/prl/abstract/10.1103/PhysRevLett.99.093902}{Phys.~Rev.~Lett. {\bf 99}, 093902 (2007).}

\bibitem{Genes2008}
C.~Genes, D.~Vitali, P.~Tombesi, S.~Gigan, and M.~Aspelmeyer, {\em Ground-state cooling of a micromechanical oscillator: Comparing cold damping and cavity-assisted cooling schemes}, \href{http://journals.aps.org/pra/abstract/10.1103/PhysRevA.77.033804}{Phys.~Rev.~A {\bf 77}, 033804 (2008).}

\bibitem{WilsonRae08}
I.~Wilson-Rae, N.~Nooshi, J.~Dobrindt, T.~J.~Kippenberg, and W.~Zwerger, {\em Cavity-assisted backaction cooling of mechanical resonators}, \href{http://iopscience.iop.org/article/10.1088/1367-2630/10/9/095007/meta}{New.~J.~Phys. {\bf 10}, 095007 (2008).}

\bibitem{MilburnWisemanBook} 
H.~M.~Wiseman and G.~J.~Milburn, {\em Quantum Measurement and Control} (Cambridge University Press, Cambridge, 2009).

\bibitem{Wiseman93}
H.~M.~Wiseman and G.~J.~Milburn, {\em Quantum theory of field-quadrature measurements}, \href{http://journals.aps.org/pra/abstract/10.1103/PhysRevA.47.642}{Phys.~Rev.~A {\bf 47}, 642 (1993).}

\bibitem{Corney98}
J.~F.~Corney and G.~J.~Milburn, {\em Homodyne measurements on a Bose-Einstein condensate}, \href{http://journals.aps.org/pra/abstract/10.1103/PhysRevA.58.2399}{Phys.~Rev.~A {\bf 58}, 2399 (1998).}

\bibitem{JacobsRev}
K.~Jacobs and D.~A.~Steck, {\em A straightforward introduction to continuous quantum measurement}, \href{http://www.tandfonline.com/doi/abs/10.1080/00107510601101934}{Contemp.~Phys. {\bf 47}, 279 (2006).}

\bibitem{Cirio2012}
M.~Cirio, G.~K.~Brennen, and J.~Twamley, {\em Quantum Magnetomechanics: Ultrahigh-Q-Levitated Mechanical Oscillators}, \href{http://journals.aps.org/prl/abstract/10.1103/PhysRevLett.109.147206}{Phys.~Rev.~Lett. {\bf 109}, 147206 (2012).}

\bibitem{Via2015}
G.~Via, G.~Kirchmair, and O.~Romero-Isart, {\em Strong Single-Photon Coupling in Superconducting Quantum Magnetomechanics}, \href{http://journals.aps.org/prl/abstract/10.1103/PhysRevLett.114.143602}{Phys.~Rev.~Lett. {\bf 114}, 143602 (2015).}

\bibitem{Minniberger2014}
S.~Minniberger, F.~Diorico, S.~Haslinger, C.~Hufnagel, C.~Novotny, N.~Lippok, J.~Majer, C.~Koller, S.~Schneider, and J.~Schmiedmayer, {\em Magnetic conveyor belt transport of ultracold atoms to a superconducting atomchip}, \href{http://link.springer.com/article/10.1007\%2Fs00340-014-5790-5}{App.~Phys.~B {\bf 116}, 1017 (2014).}

\bibitem{jackson} 
J.~D.~Jackson, {\em Classical electrodynamics (3rd Edition)} (Wiley, New York, 1962).

\bibitem{tinkham}
M.~Tinkham, {\em Introduction to superconductivity (2nd Edition)} (Dover, New York, 1996).

\bibitem{revmodern}
J.~Fortagh, and C.~Zimmermann, {\em Magnetic microtraps for ultracold atoms}, \href{http://journals.aps.org/rmp/abstract/10.1103/RevModPhys.79.235}{Rev.~Mod.~Phys. {\bf 79}, 235 (2007).}

\bibitem{Thywissen}
J.~H.~Thywissen, M.~Olshanii, G.~Zabow, M.~Drndic, K.~S.~Johnson,  R.~M.~Westervelt, and M.~Prentiss, {\em Microfabricated magnetic waveguides for neutral atoms}, \href{http://epjd.epj.org/articles/epjd/abs/1999/10/d8303/d8303.html}{Eur.~Phys.~J.~D {\bf 7}, 361 (1999).}

\bibitem{Koch2007}
J.~Koch, T.~M.~Yu, J.~Gambetta, A.~A.~Houck, D.~I.~Schuster, J.~Majer, A.~Blais, M.~H.~Devoret, S.~M.~Girvin, and R.~J.~Schoelkopf, {\em Charge-insensitive qubit design derived from the Cooper pair box}, \href{http://journals.aps.org/pra/abstract/10.1103/PhysRevA.76.042319}{Phys.~Rev.~A {\bf 76}, 042319 (2007).}

\bibitem{Doolin2014}
C.~Doolin, B.~D.~Hauer, P.~H.~Kim, A.~J.~R.~MacDonald, H.~Ramp, and J.~P.~Davis,
{\em Nonlinear optomechanics in the stationary regime}, \href{https://journals.aps.org/pra/abstract/10.1103/PhysRevA.89.053838}{Phys.~Rev.~A {\bf 89}, 053838 (2014)}

\bibitem{Praiso2015}
T.~K.~Para\"iso, M.~Kalaee, L.~Zang, H.~Pfeifer, F.~Marquardt, and O.~Painter, {\em Position-Squared Coupling in a Tunable Photonic Crystal Optomechanical Cavity}, \href{https://journals.aps.org/prx/abstract/10.1103/PhysRevX.5.041024}{Phys. Rev. X {\bf 5}, 041024, (2015)}.

\bibitem{Brawley2016}
G.~A.~Brawley, M.~R.~Vanner, P.~E.~Larsen, S.~Schmid, A.~Boisen, and W.~P.~Bowen,
{\em Non-linear optomechanical measurement of mechanical motion}, \href{http://www.nature.com/doifinder/10.1038/ncomms10988}{Nature Comm. {\bf 7}, 10988 (2016)}.

\bibitem{VenkateshPrep}
B. Prasanna Venkatesh \etal, {\em In preparation}.

\bibitem{SYB1} 
S.~Y.~Buhmann, {\em Dispersion Forces I} (Springer Tracts in Modern Physics, Berlin, 2012).

\bibitem{DIP} 
K.~Sinha \etal, In preparation. 

\bibitem{AnnettBook} 
J.~Annett, {\em Superconductivity, Superfluids and Condensates} (Oxford University Press, Oxford, 2004).

\bibitem{kagra}
C.~Tokoku \etal, {\em Cryogenic system for the interferometric cryogenic gravitationalwave telescope, KAGRA - design, fabrication, and performance test -}, \href{http://scitation.aip.org/content/aip/proceeding/aipcp/10.1063/1.4860850}{AIP~Conf.~Proc. {\bf 1573}, 1254 (2014).}

\bibitem{Skagerstam2006}
B.~K.~Skagerstam, U.~Hohenester, A.~Eiguren, and P.~K.~Rekdal, {\em Spin Decoherence in Superconducting Atom Chips}, \href{http://journals.aps.org/prl/abstract/10.1103/PhysRevLett.97.070401}{Phys.~Rev.~Lett. {\bf 97}, 070401 (2006).}

\bibitem{Nirrengarten2006}
T.~Nirrengarten, A.~Qarry, C.~Roux, A.~Emmert, G.~Nogues, M.~Brune, J.-M.~Raimond, and S.~Haroche, {\em Realization of a Superconducting Atom Chip}, \href{http://journals.aps.org/prl/abstract/10.1103/PhysRevLett.97.200405}{Phys.~Rev.~Lett. {\bf 97}, 200405 (2006).}

\bibitem{Bernon2013}
S.~Bernon \etal, {\em Manipulation and coherence of ultra-cold atoms on a superconducting atom chip}, \href{http://www.nature.com/ncomms/2013/130829/ncomms3380/full/ncomms3380.html}{Nat.~Commun. {\bf 4}, 2380 (2013).}

\bibitem{Bassi2013}
A.~Bassi, K.~Lochan, S.~Satin, T.~P.~Singh, and H.~Ulbricht, {\em
Models of wave-function collapse, underlying theories, and experimental tests}, \href{http://journals.aps.org/rmp/abstract/10.1103/RevModPhys.85.471}{Rev.~Mod.~Phys. {\bf 85}, 471 (2013).}

\bibitem{Ghirardi1986}
G.~C.~Ghirardi, A.~Rimini, and T.~Weber, {\em 
Unified dynamics for microscopic and macroscopic systems}, \href{http://journals.aps.org/prd/abstract/10.1103/PhysRevD.34.470}{Phys.~Rev.~D {\bf 34}, 470 (1986).}

\bibitem{Haslinger2013}
P.~Haslinger, N.~D\"orre, P.~Geyer,	J.~Rodewald, S.~Nimmrichter, and M.~Arndt, {\em A universal matter-wave interferometer with optical ionization gratings in the time domain}, \href{http://www.nature.com/nphys/journal/v9/n3/abs/nphys2542.html}{Nat.~Phys. {\bf 9}, 144 (2013).}



\end{thebibliography}
\end{document}